\documentclass[aps,pra,superscriptaddress,amsmath,amssymb,showpacs,twocolumn,notitlepage]{revtex4-1}
\usepackage{amssymb}
\usepackage{amsmath}
\usepackage{graphicx}
\usepackage{bm}
\usepackage{bbm}
\usepackage{color}
\usepackage{xcolor}
\usepackage{subfigure}

\newcommand{\Q}{\dot{Q}}

\graphicspath{{./Figures/}}

\begin{document}

\title{Quantum thermal machine acting on a many-body quantum system:\\role of correlations in thermodynamic tasks}

\author{Pierre Doyeux}
\affiliation{Laboratoire Charles Coulomb (L2C), UMR 5221 CNRS-Universit\'{e} de Montpellier, F- 34095 Montpellier, France}

\author{Bruno Leggio}
\affiliation{Laboratoire Charles Coulomb (L2C), UMR 5221 CNRS-Universit\'{e} de Montpellier, F- 34095 Montpellier, France}

\author{Riccardo Messina}
\affiliation{Laboratoire Charles Coulomb (L2C), UMR 5221 CNRS-Universit\'{e} de Montpellier, F- 34095 Montpellier, France}

\author{Mauro Antezza}
\affiliation{Laboratoire Charles Coulomb (L2C), UMR 5221 CNRS-Universit\'{e} de Montpellier, F- 34095 Montpellier, France}
\affiliation{Institut Universitaire de France, 1 rue Descartes, F-75231 Paris Cedex 05, France}

\begin{abstract}
We study the functioning of a three-level thermal machine when acting on a many-qubit system, the entire system being placed in an electromagnetic field in a stationary out-of-thermal-equilibrium configuration. This realistic setup stands in between the two so-far explored cases of single-qubit and macroscopic object targets, providing information on the scaling with system size of purely quantum properties in thermodynamic contexts. We show that, thanks to the presence of robust correlations among the qubits induced by the field, thermodynamic tasks can be delivered by the machine both locally to each qubit and collectively to the many-qubit system: this allows a task to be delivered also on systems much bigger than the machine size.
\end{abstract}


\maketitle

\section{Introduction}
The study and the exploitation of out-of-equilibrium quantum properties at micro and nanoscale, and at the level of few body systems, are becoming more and more important in pure and applied research \cite{Haenggi2009,Li2012,Esposito2009,Abah2012,Leggio2013a,Leggio2013b,Gour2015,Scovil1959,Linden2010,Skrzypczyk2011,Venturelli2013,Bellomo2012a,Bellomo2012b,Esposito2010,Leggio2015a,Leggio2015b,Leggio2015c}. Among the topics recently attracting a great deal of attention, a particular mention deserves the so-called branch of quantum thermodynamics \cite{GemmerBook, Horodecki2013, Skrzypczyk2014, Binder2015}. As much as its classical counterpart, indeed, its consequences bear great theoretical, experimental and technological importance.

Central topic of thermodynamics, the concept of thermal machine well represents this multidisciplinary spirit by connecting profound theoretical ideas (such as the notions of entropy and irreversibility) to direct applicative outcomes. In the same framework, the fast-paced development of the idea of quantum thermal machine \cite{Scovil1959,Linden2010,Venturelli2013,Levy2012,Fialko2012,Klimovsky2013,Correa2014,Correa2013} provided in these latest years an ideal scenario to explore the possible practical implications of purely quantum features \cite{Brunner2014,Leggio2015a,Mitchison2015,Brask2015} as, e.g., quantum coherence between single quantum emitters (hereby referred to as \textit{atoms}).

Among the possible models of thermal machines available from classical contexts, remarkable importance in quantum scenarios has been given to the so-called absorption (or self-contained) machines \cite{Scovil1959,Linden2010,Venturelli2013,Levy2012,Correa2014,Leggio2015a}. These systems can indeed deliver thermodynamic tasks without the need of external work supplies, avoiding the problem of addressing and controlling single quantum systems.

The prototype of absorption quantum thermal machine is nowadays a few-level atom interacting with a target body, on which the thermodynamic task (refrigeration, heating up, work) has to be delivered. Two different limiting cases have so-far been explored: on the one hand, many have studied the situation in which the target body is a macroscopic system at a certain temperature \cite{Levy2012,Correa2014,Correa2013,Correa2014b}, the task thus being a stationary heat flux produced by the machine from/into the target system. In this first case, despite being the setup of applicative interest, the quantum features of machine and machine-target interaction are suppressed and their role on the task becomes mostly irrelevant \cite{Correa2013}; the opposite limiting case, also often studied, is the action of the machine on a single quantum system, mostly in its simplest form represented as a two-level atom (or qubit) \cite{Linden2010,Skrzypczyk2011,Leggio2015a,Brunner2014}. Although simplified, these models allow to directly highlight the role of quantum properties (quantum coherence, quantum discord, entanglement) in the machine-target interaction and in the final delivery of the task.
A gap thus persists in the understanding of quantum thermal machines. In particular, the questions of how the quantum properties and their role in thermodynamic tasks scale with the size of the target system, and whether they can represent a resource the machine can use to act on bigger and bigger (quantum) systems, remain unaddressed.

\begin{figure}[h!]
\begin{center}
\includegraphics[width=245pt]{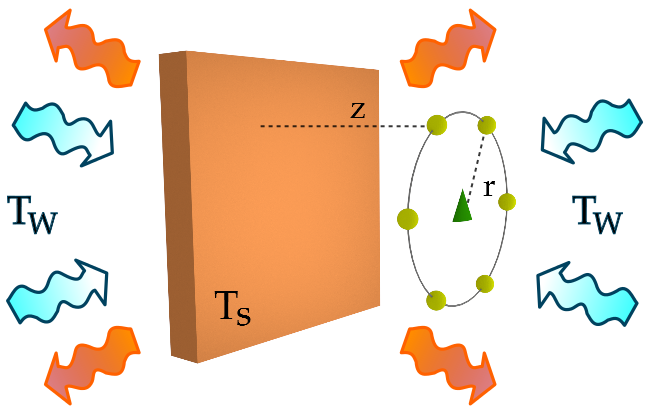}
\end{center}
\caption{The physical setup consists of an OTE electromagnetic field produced by a sapphire slab of thickness $\delta$ and temperature $T_S$, embedded in a thermal blackbody radiation at temperature $T_W\neq T_S$. Such a field plays the role of Markovian environment for a system of quantum emitters (atoms), all placed at the same distance $z$ from the slab surface. Four of these atoms are qubits, placed in a regular disposition along a circle of radius $r$, the center of which is occupied by a three-level atom. The qubit system is the target body, on which the three-level machine M delivers thermodynamic tasks. For each plot of this paper, the radius is fixed at $r=0.833\, \mu$m, unless otherwise specified.}
\label{Figure1}
\end{figure}

This paper is a first step in filling this gap. Here we study the thermodynamics of a system composed of a three-level quantum thermal machine resonantly coupled to $n_{\mathrm{q}}$ identical qubits. This composite atomic system is embedded in an out-of-thermal-equilibrium (OTE) electromagnetic field produced by macroscopic objects kept at different temperatures, analogously to the setup studied in \cite{Leggio2015a} in the case of $n_{\mathrm{q}}=1$. Such a field naturally couples resonant transitions, allowing both the interaction of the machine with each of the qubits and the establishment of qubit-qubit correlations in the target system. Thus, classical and quantum correlations are built at stationarity and their interplay fundamentally affects the thermodynamic properties of the qubits, which are studied both for an exemplary case at fixed qubits number and as a function of $n_{\mathrm{q}}$.

This paper is structured as follows: in Sec.~\ref{physyst} we introduce the setup of both field and atoms and describe their interaction and the consequent atomic dynamics and stationarity. The thermodynamics of the machine-target interaction is in particular analyzed in Sec.~\ref{quanttherm}. In Section \ref{Sym_case}, we investigate, as an exemplary case, the thermodynamics of the system when the target body is composed of four qubits. Sec.~\ref{scaling} is dedicated to the scaling with the number of qubits of some interesting quantities introduced in Secs.~\ref{quanttherm} and \ref{Sym_case}. Final remarks and conclusions are drawn in Sec.~\ref{concl}. Finally, technical details about the atomic master equation and all the correlation quantifiers employed in the text can be found, respectively, in Appendix \ref{rates} and \ref{correlations}.

\section{Physical system}\label{physyst}
The system we consider here, as depicted in Fig.~\ref{Figure1}, consists of a multipartite quantum system embedded in an OTE electromagnetic field. This field is produced by a macroscopic object, i.e., a sapphire slab of thickness $\delta=0.05\, \mu$m and of resonance frequency $\omega_S=0.81\times10^{14}$ rad.$\mathrm{s}^{-1}$ \cite{OpticalConstants}, kept at fixed temperature $T_S$ and placed in a region of space where a thermal blackbody radiation exists, emitted by some far-away walls at fixed temperature $T_W\neq T_S$.

At stationarity, the non-thermal electromagnetic field filling the space between slab and walls can be precisely characterized in terms of its correlation functions \cite{Messina2011a,Messina2011b,Messina2014}. For a detailed description of its properties, we refer the interested reader to \cite{Bellomo2012a,Bellomo2012b,Messina2011a,Messina2011b,Messina2014}. In this region of space, at a distance $z$ from the slab surface, a multipartite quantum system is placed, consisting of $n_{\mathrm{q}}$ identical qubits (the target body B) of frequency $\omega_{\mathrm{q}}=0.1 \times \omega_S$, placed on a circle at the center of which a three level atom (the machine M) lies.  The circle is parallel to the slab surface, so that $z$ is the same for every atom. Finally, the radius of the circle is referred to as $r$.

M has 3 allowed transitions between its three levels $|0\rangle, |1\rangle$ and $|2\rangle$. The transition between $|1\rangle$ and $|2\rangle$, labeled as 2, has the same frequency of the qubits ($\omega_2=\omega_{\mathrm{q}}$), whereas the other two satisfy $\omega_1\neq\omega_2$ and $\omega_3=\omega_1+\omega_2$. Besides, $\omega_3=\omega_S$ such that the corresponding transition $|0\rangle\leftrightarrow|2\rangle$ is much more affected by the slab than the other two \cite{Bellomo2012b}. The Hamiltonian of the total field+atoms system then reads
\begin{equation}
H_{\mathrm{tot}}=H_{\mathrm{emitters}}+H_{\mathrm{field}}+H_I,
\end{equation}
in terms of the free emitters and field Hamiltonians $H_{\mathrm{emitters}}$ and $H_{\mathrm{field}}$ and the atoms-field interaction Hamiltonian $H_I$. When expressed under the dipole-approximation limit \cite{CohenTannBook}, $H_I=-\sum_{i,n}\mathbf{d}^{(n)}_i \cdot \mathbf{E}(\mathbf{R}_n)$. In the absence of permanent atomic dipoles, $\mathbf{d}^{(n)}_i$ is the field-induced dipole moment of the $i$-th transition of the atom $n$ which is located at $\mathbf{R}_n$. The electromagnetic field at this position is $\mathbf{E}(\mathbf{R}_n)$.
\subsection{The master equation}
In the weak atom-field coupling limit and under the rotating wave approximation, a Markovian master equation \cite{BreuerBook} for the atomic density matrix $\rho$ can be given \cite{Bellomo2012a,Bellomo2012b,Bellomo2013a,Bellomo2013b} under the form
\begin{equation}\label{METQ}
\frac{d\rho}{d t}=-\frac{i}{\hbar}\big[H_{\mathrm{sys}},\rho\big]+D_B(\rho)+D_M(\rho)+D_{\mathrm{nl}}(\rho),
\end{equation}
where $H_{\mathrm{sys}}=H_{\mathrm{emitters}}+H_{\Lambda}$ represents an effective Hamiltonian of the atomic system, in which the dipole-dipole interaction term
\begin{equation}\label{HL}
H_{\Lambda}=\sum_{n=1}^{n_{\mathrm{q}}}\hbar\Lambda_{n\textrm{M}}(\sigma^{\dag}_n\kappa_2+\sigma_n \kappa_2^{\dag})+\sum_{n\neq m}^{n_{\mathrm{q}}}\hbar\Lambda_{nm}\sigma^{\dag}_n\sigma_m
\end{equation}
has been added to the free atomic Hamiltonian, where $\sigma_n$ is the lowering operator of the $n$-th qubit and $\kappa_t$ is the lowering operator corresponding to the $t$-th transition of the machine M. This interaction couples only resonant transitions in the atomic system, and allows qubits and machine to coherently exchange excitations. It is worth stressing at this point that we assume that the physics of our system is robust with respect to a small dephasing, as it is the case in many analogous studies \cite{Feist}.

It is important to stress here that the dipole-dipole interaction amplitudes $\Lambda_{n\textrm{M}}$ and $\Lambda_{nm}$ crucially depend, for each pair of atoms, on the mutual orientation of the two dipoles. In particular, consider a generic pair $(n,m)$ of atoms (which, possibly, can also include the machine) in a plane parallel to the slab, and let the $x$ axis be the direction of the line joining the two atoms. The resonant dipole-dipole interaction between these two atoms has then only components $x-x$, $y-y$, $z-z$ and $x-z$. All these components have both a contribution from the free field (in the absence of the slab) and a reflected contribution due to the scattering properties of the slab (see Eq.~\eqref{lambda0R} and \cite{Bellomo2013a,Bellomo2013b} for all the technical details), with the only exception of the $x-z$ component, whose only contribution stems from the reflected field. As a consequence, $x-z$ interactions are weaker than the other components. It is worth stressing that the reference frame used here changes each time a new pair is chosen and must then be carefully set before starting to calculate the coefficients $\Lambda$.

The terms $D_\mathrm{B}(\rho)=\sum_{n=1}^{n_{\mathrm{q}}}D_\mathrm{B}^{(n)}(\rho),\,\,\,D_\mathrm{M}(\rho)=\sum_{t=1}^3D_\mathrm{M}^{(t)}(\rho)$ and $D_{\mathrm{nl}}(\rho)=\sum_{n=1}^{n_{\mathrm{q}}}D_{\mathrm{nl}}^{(n\mathrm{M})}(\rho)+\sum_{n\neq m}^{n_{\mathrm{q}}}D_{\mathrm{nl}}^{(nm)}(\rho)$ describe dissipative effects in the atomic dynamics, induced by the interaction with the OTE field. They are
\begin{eqnarray}
D_\mathrm{B}^{(n)}(\rho)&=&\Gamma_n^+(\omega_\mathrm{q})\Big(\sigma_n\rho\sigma_n^{\dag}-\frac{1}{2}\big\{\sigma_n^{\dag}\sigma_n,\rho\big\}\Big)\nonumber \\
&+&\Gamma_n^-(\omega_\mathrm{q})\Big(\sigma_n^{\dag}\rho\sigma_n-\frac{1}{2}\big\{\sigma_n\sigma_n^{\dag},\rho\big\}\Big),\label{DQ}
\end{eqnarray}
representing the single qubit dissipative energy exchange with the field,
\begin{eqnarray}
D_\mathrm{M}^{(t)}(\rho)&=&\Gamma_{\mathrm{M}}^+(\omega_t)\Big(\kappa_t\rho\kappa_t^{\dag}-\frac{1}{2}\big\{\kappa_t^{\dag}\kappa_t,\rho\big\}\Big)\nonumber \\
&+&\Gamma_{\mathrm{M}}^-(\omega_t)\Big(\kappa_t^{\dag}\rho\kappa_t-\frac{1}{2}\big\{\kappa_t\kappa_t^{\dag},\rho\big\}\Big),\label{DT}
\end{eqnarray}
being the machine-field dissipative energy exchange through the $t$-th machine transition and finally
\begin{eqnarray}
D_{\mathrm{nl}}^{(n\mathrm{M})}(\rho)&=&\Gamma_{n\mathrm{M}}^+(\omega_\mathrm{q})\Big(\kappa_2\rho\sigma_n^{\dag}-\frac{1}{2}\big\{\sigma_n^{\dag}\kappa_2,\rho\big\}\Big)\nonumber \\
&+&\Gamma_{n\mathrm{M}}^-(\omega_\mathrm{q})\Big(\kappa_2^{\dag}\rho\sigma_n-\frac{1}{2}\big\{\sigma_n\kappa_2^{\dag},\rho\big\}\Big)+\mathrm{h.c.},\label{DTQM}\\
D_{\mathrm{nl}}^{(nm)}(\rho)&=&\Gamma_{nm}^+(\omega_\mathrm{q})\Big(\sigma_m\rho\sigma_n^{\dag}-\frac{1}{2}\big\{\sigma_n^{\dag}\sigma_m,\rho\big\}\Big)\nonumber \\
&+&\Gamma_{nm}^-(\omega_\mathrm{q})\Big(\sigma_m^{\dag}\rho\sigma_n-\frac{1}{2}\big\{\sigma_n\sigma_m^{\dag},\rho\big\}\Big)\label{DTQQ}
\end{eqnarray}
are non-local dissipative terms describing energy exchanges between the field and any two-atom pair in the open system. In these terms, the two atoms behave collectively and emit or absorb photons as a single entity. This can be shown by noting that the heat flux in/out each two-atom pair is proportional to the coherence in the reduced two-atom system, indicating that such emission/absorption processes are due to the correlations between the atoms. Specifically, as we will show later on, the change in internal energy of each of the two atoms in a pair, due to such non-local dissipation, is exactly the same.

Note that $\Gamma_{nm}^{\pm}$ and $\Gamma_{n\mathrm{M}}^{\pm}$ can be decomposed in contributions related to dipole components along the line joining the two atoms, perpendicular to it on the $xy$ plane and perpendicular to the slab in exactly the same way as done for $\Lambda$. The only non-zero contributions are also in this case $x-x$, $y-y$, $z-z$ and $x-z$. This is a manifestation of the fact that non-local dissipation and dipole-dipole coupling are related, respectively, to the imaginary and to the real part of the electromagnetic field Green function at two different points in space.

All the relaxation rates $\Gamma_n^{\pm}$, $\Gamma_{\mathrm{M}}^{\pm}$, $\Gamma_{n\mathrm{M}}^{\pm}$ and $\Gamma_{nm}^{\pm}$ and the dipole-dipole interaction strength $\Lambda$ depend on the frequency of the associated transition, on the ground-excited states matrix element $\mathbf{d}$ of the dipole operator of the transition, on the two externally-fixed temperatures $T_S$ and $T_W$ and on the material properties of the slab. The detailed expressions for all these parameters can be found in \cite{Bellomo2013b} and are given in the Appendix \ref{rates}.\\

\subsection{Quantum thermodynamics\\of the system}\label{quanttherm}
In what follows, we will employ different quantities describing the thermodynamics of the field+atoms system. In particular, two classes of parameters will stand out for their importance in our study: temperatures and heat fluxes. The definitions and classification we will use throughout this paper strictly follows the ones given in \cite{Leggio2015a}.

The definition of heat fluxes in Markovian frameworks goes through the first law for quantum systems \cite{BreuerBook,Wichterich2007}. Its form is easily given as the time variation of the mean value of their Hamiltonian, which represents in quantum contexts the internal energy of a system. The same fluxes play a major role in the more delicate generalization of the second law, discussed in Appendix \ref{App2L}).

Given the fact that the unitary term in \eqref{METQ} commutes with the Hamiltonian of each atom, and thus also with the total Hamiltonian of the atomic system, and there being by construction no external work in our system (such that $\partial H/\partial t=0$), the only possibility for the change in internal energy $U=\langle H\rangle$ of an atom or a collection of atoms is given by heat fluxes. Note that, seen by a subset of atoms, also the (global) unitary term $-\frac{i}{\hbar}\big[H_{\mathrm{sys}},\rho\big]$ can produce a change in the internal energy and in the entropy of the subset. Each dissipative process $D$ produces a change in $U$ given by
\begin{equation}\label{heatflux}
\dot{U}_D=\mathrm{tr}\left(H D(\rho)\right)=\dot{Q}_D,
\end{equation}
$\rho$ being the state of the atomic system at the time instant of interest. Eq.~\eqref{heatflux} is the definition of the heat flux generated in the system with Hamiltonian $H$ due to the dissipative process $D$.

Temperature, on the other hand, is a tricky quantity to define in systems far from their thermodynamic limit. The best one can do is to recur to some analogy with known properties of temperature in macroscopic classical systems. The property we turn to for the characterization of our system is that a temperature gradient between two bodies imposes a direction to the heat flux between them.

The (effective) temperature of the OTE field is well defined in terms of the two real temperatures $T_W$ and $T_S$ and the slab material. The temperature of a thermal field can be inferred from the photon emission/absorption rates of an atomic transition interacting with it, independently on the transition frequency. In the case of the OTE field considered here, however, different transitions naturally ``feel" different field temperature (or, in other words, the ratio of emission to absorption rates is not simply an exponential function of the transition frequency). This effective environmental temperature felt by the $i$-th atomic transition of frequency $\omega_i$ can be defined as
\begin{equation}\label{Tn}
T_i=\frac{\hbar \omega_i}{k_B \ln(\Gamma^+(\omega_i)/\Gamma^-(\omega_i))},
\end{equation}
where the $\Gamma^{\pm}(\omega_i)$ are the single transition dissipative rates involved in the master equation \eqref{METQ}, whose explicit expression can be found in Eqs.~\eqref{locgp}-\eqref{locgm} in Appendix \ref{rates}. This environmental temperature describes the way the OTE field exchanges heat with any two-level object having a transition frequency $\omega_i$.

Having now at disposal both the expression of the field temperature and of the heat flux between the field and an atomic transition, one can identify an equivalent parameter describing the way the transition exchanges heat with the field. Indeed, employing Eq.~\eqref{heatflux} to calculate the local heat flux produced by the local dissipative process in either Eq.~\eqref{DQ} (for each qubit) or Eq.~\eqref{DT} (for each machine transition) one obtains
\begin{equation}\label{fluxtemp}
\dot{Q}_i=X_i\left(e^{\frac{\hbar \omega_i}{k_B \theta_i}}-e^{\frac{\hbar \omega_i}{k_B T_i}}\right),
\end{equation}
where $X_i>0$ and
\begin{equation}\label{thetai}
\theta_i=\frac{\hbar \omega_i}{k_B \ln(p^g_i/p^e_i)},
\end{equation}
having introduced the ground (excited) state of the $i$-th transition $p^g_i$ ($p^e_i$). As one easily sees, $\theta_i$ (hereby referred to as the population temperature) plays here the role of temperature for atomic transition, as now $T_i$ and $\theta_i$ characterize the heat exchanged by the transition with the external field in a symmetric way. Moreover, the heat matches now the requirement to flow from the hotter into the colder object.

Aside of the heat exchanged locally between each atom and the field, two other fluxes affect the internal energy of atoms, stemming respectively from the atom-atom dipolar coupling $\Lambda$ in Eq.~\eqref{HL} (resonant heat flux $\Q_r$) and from the collective nonlocal dissipation in Eqs.~\eqref{DTQM} and \eqref{DTQQ} (nonlocal heat flux $\Q_d$). Both of these fluxes couple only resonant atomic transitions. Whereas the first flux does not change the total energy of the atomic system, representing a hopping of excitations from one atom into another one, the nonlocal flux $\Q_d$ implies a net flux going in/out the atomic system and being sustained by the environment: due to such nonlocal dissipation terms, the presence of an atom triggers collective emission or absorption of photons with any other atom being in resonance with it. Seen from the point of view of the internal energy of each atom in the pair, this phenomena produce heat fluxes with the environment, caused by the presence of a second atom.

Straightforward specialization of the definition \eqref{heatflux} for either dipole-dipole coupling or collective dissipation for a pair $(n,m)$ of atoms (possibly including the machine M) gives
\begin{eqnarray}
\Q_r(m\rightarrow n)&=&2\hbar\omega_\mu\Lambda_{nm}\mathrm{Im}[c_{nm}],\label{qr}\\
\Q_d(n,m)&=&-\hbar\omega_\mu\mathrm{Re}\left[c_{nm}\big(\Gamma_{nm}^+-(\Gamma_{nm}^-)^*\big)\right]\label{qd},
\end{eqnarray}
where $c_{nm}$ is the coherence in the reduced two-atom state $\rho_{nm}=\mathrm{tr}_{p\neq n,m}(\rho)$, when expressed in the ordered basis $\{|g_ng_m\rangle,|g_ne_m\rangle,|e_ng_m\rangle,|e_ne_m\rangle\}$, $|g_n\rangle$ ($|e_n\rangle$) being the ground (excited) state of the transition of atom $n$. Note that $\Q_r(m\rightarrow n)$ is the flux flowing from $m$ to $n$, meaning that $\Q_r(m\rightarrow n)>0$ represents energy going out of atom $m$ and into atom $n$. On the other hand, $\Q_d(n,m)$ has the same sign for both atoms: $\Q_d(n,m)>0$ means that both $n$ and $m$ are absorbing photons from the field.

Through these two heat fluxes, atoms can exchange energy and, in particular, the machine can deliver thermodynamic tasks on the target qubit system. As comes clear from Eqs.~\eqref{qr} and \eqref{qd}, the thermodynamics of the machine functioning is based on the presence of quantum coherence between the machine and its target body.

These energy fluxes will have the effect of changing the qubit population temperatures with respect to their corresponding environmental temperatures: Eqs.~\eqref{Tn} and \eqref{thetai} are the main quantities we will study for our system. In particular, being $T_i$ the temperature at which each qubit would thermalize in absence of the rest of the atomic system, we will define a thermodynamic task as a stationary modification of the qubit temperature $\theta_i$ with respect to the corresponding value of $T_i$.

Previous works on this model \cite{Leggio2015a} have shown that the machine is able to deliver different tasks when interacting with a single qubit. In particular, under certain conditions, qubit population inversion can be achieved. For this reason throughout this paper, for graphical and technical purposes, we will work with the parameter $-\beta=-\theta^{-1}$, which is an increasing function of $\theta$ and avoids the divergent behavior shown by the temperature in correspondence to a point of population inversion.
\section{4-qubit symmetric configuration}
\label{Sym_case}
\begin{figure}[t!]
\begin{center}
\includegraphics[scale=0.18]{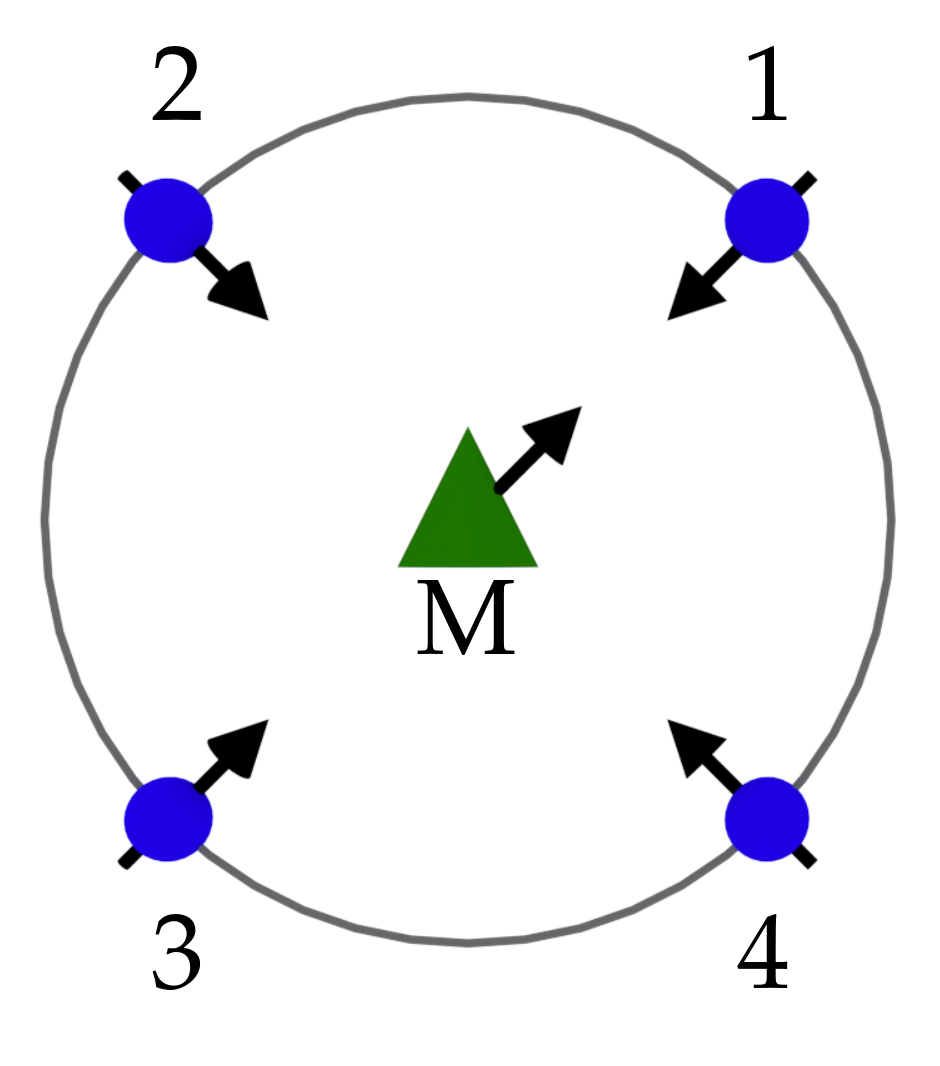}
\end{center}
\caption{Geometric configuration of the atomic sytem. The qubits are regularly distributed along a circle centered on the machine $M$. Every atom has the same $z$. The dipole of each qubit points toward the machine, whereas the machine's one points toward qubit 1. The interaction between two atoms depends on the projection of their dipoles along the axis joining them. For example, the qubit $2$ does not interact with $M$ but interacts with $1$.}
\label{Figure2}
\end{figure}

As an exemplary case, we study the symmetric configuration represented in Fig.~\ref{Figure2}, where four qubits are regularly distributed on a circle centered on the machine and parallel to the slab.

\begin{figure}[t!]
\begin{center}
\includegraphics[width=245pt]{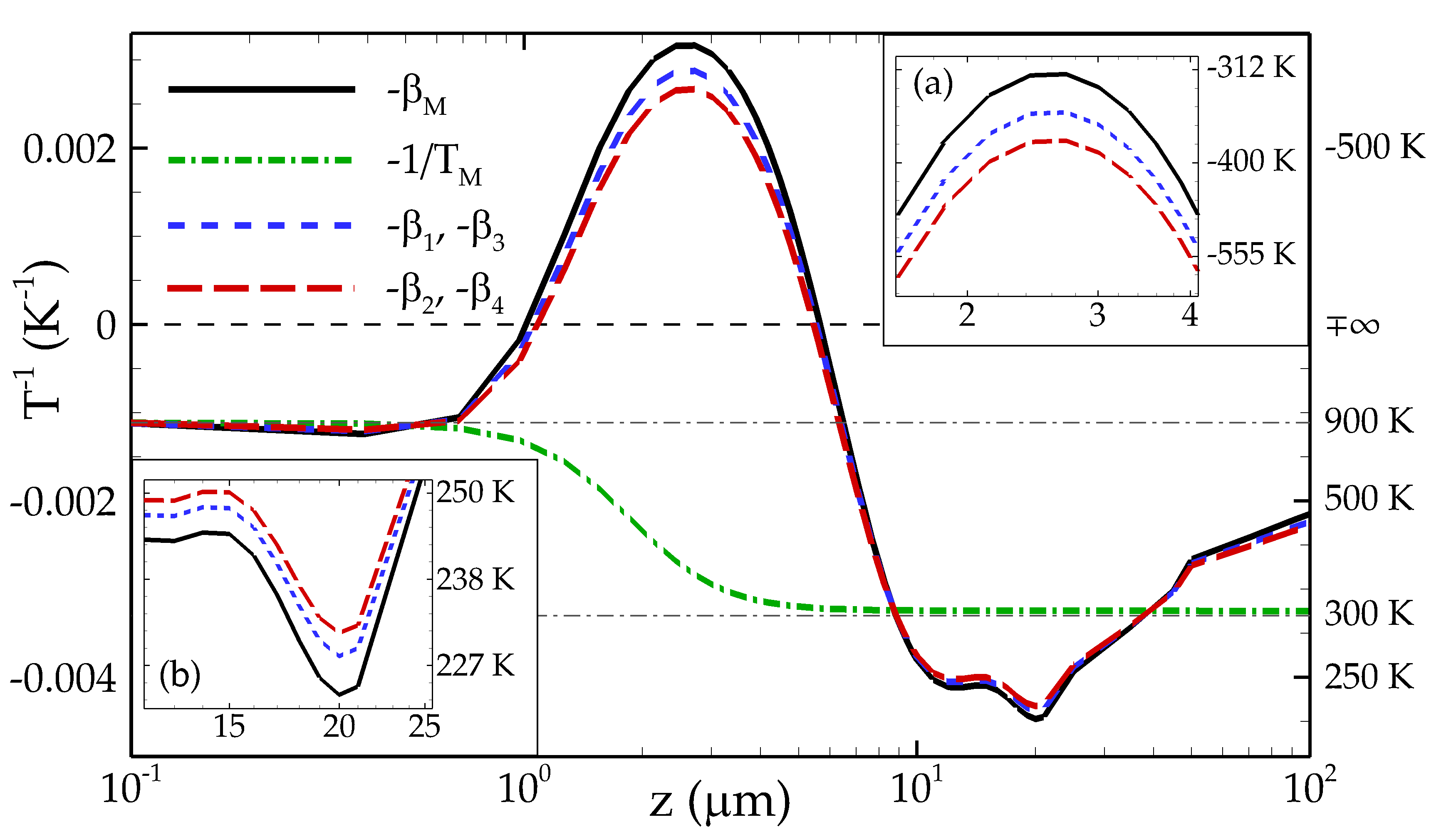}
\end{center}
\caption{
Left vertical scale: $-\beta_M$ (solide black line), $-1/T_M$ of the resonant transition of the machine (dot-dashed green line), $-\beta_1=-\beta_3$ (short-dashed blue line),   $-\beta_2=-\beta_4$ (long-dashed red line) versus the slab-atom distance $z$ for the configuration of Fig.~\ref{Figure2}. Right vertical scale: temperatures in correspondence with the left scale, externally fixed temperatures $T_W=300\, $K and $T_S=900\, $K (grey dot-dashed lines). The equalities $-\beta_1=-\beta_3$ and $-\beta_2=-\beta_4$ are due to the symmetry of the system. Panels ($a$) and ($b$) show the extremum of heating ($a$)  and cooling ($b$). Notice that in the heating region, the population temperatures can reach negative values ($-\beta$ positive) meaning that the qubits undergo population inversion.}
\label{Figure3}
\end{figure}

This means that every atom has the same $z$. In this case, the dipole of each qubit is pointing toward the machine (labeled as M) whose dipole points toward one of the qubits which we label as $1$. The rest of the qubits is indexed from $2$ to $4$ in the counterclockwise direction. We begin the analysis of this system with Fig.~\ref{Figure3} where the inverse of both environmental and population temperatures of the resonant transition of the machine, as well as the inverse of the population temperature of each qubit are plotted versus $z$.

Since every dipole is parallel to the slab, the environmental temperature is the same for every qubit and also for the resonant transition of the machine. For small values of $z$, the environmental electromagnetic field is mainly affected by the contribution of the slab. In this situation $T_{\mathrm{M}}$ in then extremely close to $T_S$. On the contrary, for very large $z$, the contribution of the walls to the environmental electromagnetic field is dominant, thus $T_{\mathrm{M}}$ gets close to $T_W$. For intermediate values of $z$, the environmental temperature as defined through the transition rates in \eqref{Tn} has intermediate values in $[T_W,T_S]$. These rates depend on several parameters such as $z$, $T_S$, $T_W$, the slab dielectric properties and its thickness (see Appendix \ref{rates}).

Due to the OTE configuration, the environmental temperatures of the machine can be different from their population ones. In particular, for its resonant transition one has $T_{\mathrm{M}} \neq \theta_{\mathrm{M}}$. This is due to the fact that each transition of M feels a different environmental temperature, which in turn depends on $z$. Thus a change in $z$ modifies the populations distribution of M and, as a consequence, tunes $\theta_{\mathrm{M}}$ (for more details see \cite{Bellomo2012a,Bellomo2012b,Leggio2015a}). Note that, unlike $T_{\mathrm{M}}$, $\theta_{\mathrm{M}}$ reaches higher (lower) temperatures than the highest (lowest) temperature externally fixed ($T_W$ and $T_S$). Notably, $\theta_{\mathrm{M}}$ can also be brought to negative values, meaning that the resonant transition of M is in population inversion.

Similarly to \cite{Leggio2015a}, where the machine is acting on a single qubit, M delivers thermodynamic tasks on qubits $1$ and $3$. These qubits are indeed the only ones interacting both with their local environment and with the resonant transition of M. As such, they reach a steady temperature which is in between $T_{\mathrm{M}}$ and $\theta_{\mathrm{M}}$. In particular, due to the strong coupling with the machine ($\Lambda_{{\mathrm{M}}1({\mathrm{M}}3)}>>\Gamma^{\pm}_{1(3)}$), $\theta_{1(3)}$ will be much closer to $\theta_{\mathrm{M}}$ than to $T_{\mathrm{M}}$. Remarkably, also in this configuration M can perform strong heating or cooling: $\theta_{1(3)}$ can indeed be brought to values outside the range $[T_W,T_S]$ and, in particular, to negative values (population inversion). These interactions are notably due to the fact that the dipoles of M, qubit $1$ and qubit $3$ are collinear. However, the dipoles of $2$ and $4$ are orthogonal to the one of the machine, therefore M is not coupled to them. Yet, as one can see in Fig.~ \ref{Figure3}, qubits $2$ and $4$ undergo the same thermodynamic tasks as $1$ and $3$.

Indeed, even though there is no $x-x$ or $y-y$ interaction for the pairs $\{{\mathrm{M}},2\}$ and $\{{\mathrm{M}},4\}$, this is not the case for $\{1,2\}$ ($\{1,4\}$) and $\{3,2\}$ ($\{3,4\}$), thus inducing non-zero interactions between all the qubits. Therefore, similarly to the task undergone by $1$ ($3$), the population temperature of qubit $2$ ($4$) reaches a steady temperature $\theta_{2(4)} \in [T_{\mathrm{M}},\theta_1]$. In other words, qubits $1$ and $3$ relay the tasks delivered on them by M to qubits $2$ and $4$, despite these latter ones have no direct interaction with the machine.

As just discussed, the machine can heat up or cool down qubits that are not necessarily coupled to it thanks to qubit-qubit interactions. A way of understanding how subparts of a quantum system interact with each other is to look into their correlations.

\begin{figure}[h!]
\begin{center}
\includegraphics[width=245pt]{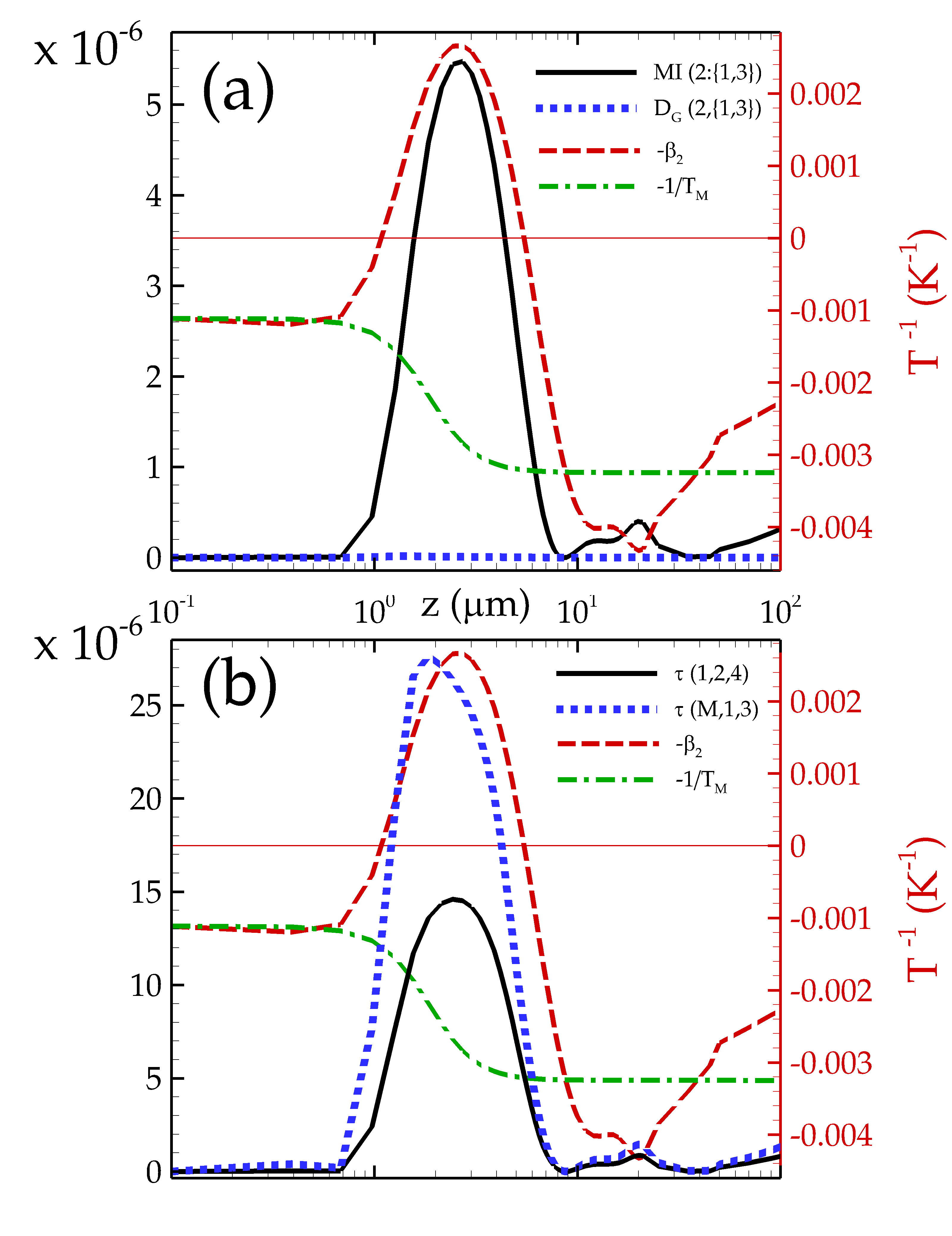}
\end{center}
\caption{Panel ($a$): Left vertical scale: mutual information MI (solid black line) and geometric quantum discord $D_\mathrm{G}$ (short-dashed blue line) of the bipartition ($2$,$\{1,3\}$) in the qubits system versus $z$ (slab-atoms distance). Panel ($b$): Left vertical scale : tripartite correlations for the tripartitions $(1,2,4)$ (solid black line) and $(M,1,3)$ (blue short-dashed line) versus $z$. The right vertical scale of both panels shows the values of inverse of the population temperature $-\beta_2$ of qubit $2$ (long-dashed red line) and the inverse of the atomic environmental temperature $-1/T_{\mathrm{M}}$ (dot-dashed green line).}
\label{Figure4}
\end{figure}

First, in Fig.~\ref{Figure4}a we consider the bipartite mutual information MI \cite{Cerf1997,Vedral2002} (Appenxix \ref{correlations}1). MI quantifies the total correlations between two subparts of a system. To bring out the essential role of qubits $1$ and $3$ in the thermodynamic tasks undergone by $2$, we plot the mutual information along the bipartition $(2,\{1,3\})$ versus $z$ (solid black line of Fig.\ref{Figure4}a). As one can see, MI is zero if and only if no task is achieved (i.e., when $\theta_2=T_2$), whereas the changes of $-\beta_2$ correspond to the ones of $\text{MI}(2\mathrm{:}\{1,3\})$. In particular, the two local maxima of $\text{MI}(2\mathrm{:}\{1,3\})$ are reached in correspondence to the peak in refrigeration and population inversion induced by M.

The bipartite correlations quantified by MI make no distinction between classical and quantum ones. One might wonder whether the correlations $\text{MI}(2\mathrm{:}\{1,3\})$ are of classical or quantum nature. To answer this question, we employ the quantity known as geometrical quantum discord $D_\mathrm{G}$ \cite{Modi2012,Paula2014} with the expression given in \cite{Spehner2014}, which quantifies purely quantum correlations in bipartite systems. In particular, $D_\mathrm{G}$ measures the distance in the state space between the bipartite state under investigation and the closest classical state (Appendix \ref{correlations}2). From Fig.~\ref{Figure4}a, it is clear that $D_\mathrm{G}(2,\{1,3\})$ is almost constantly zero, thus implying that the correlations between $2$ and $\{1,3\}$ are mostly of classical nature. Note however that the correlations between the machine and qubits $1$ and $3$ (not plotted) show a non-negligible quantum contribution \cite{Leggio2015a}.

Another quantifier supplying an important piece of information about correlations in this many-body quantum system is the tripartite mutual information $\tau$ \cite{Giorgi2011,Maziero2012} (Appendix \ref{correlations}3). It measures the total correlations in a tripartite system that cannot be expressed as a combination of bipartite correlations in any of its subsystems. In other words, $\tau$ characterizes the total genuinely tripartite correlations. Fig.~\ref{Figure4}b shows $\tau$ for two of the subsystems mainly involved in the two-step delivery of thermodynamic task previously described: the subsystem $\{\mathrm{M},$1$,$3$\}$, where the task is exerted by the machine on the qubits system, and the subsystem $\{1,2,4\}$ where such an effect is passed on by qubit $1$ to $2$ and $4$.

One notices two interesting features. First of all, $-\beta_2$ reaches its maximum at the same $z$ as $\tau(1,2,4)$, showing how the steady temperature distribution is ultimately due to qubit-qubit correlations. Secondly, it shows that the two stages of the task (M $\rightarrow$ $\{1,3\}$ and $\{1,3\} \rightarrow \{2,4\}$) imply a comparable amount of tripartite correlations: the maximum of $\tau(M,1,3)$ is indeed around twice as high as $\tau(1,2,4)$. Given the fact that, due to symmetry, $\tau(1,2,4)=\tau(3,2,4)$, one concludes that $\tau(\mathrm{M},1,3) \simeq \tau(1,2,4)+\tau(3,2,4)$, which allows an optimal distribution of the task among all the qubits.

\begin{figure}[t!]
\begin{center}
\includegraphics[width=245pt]{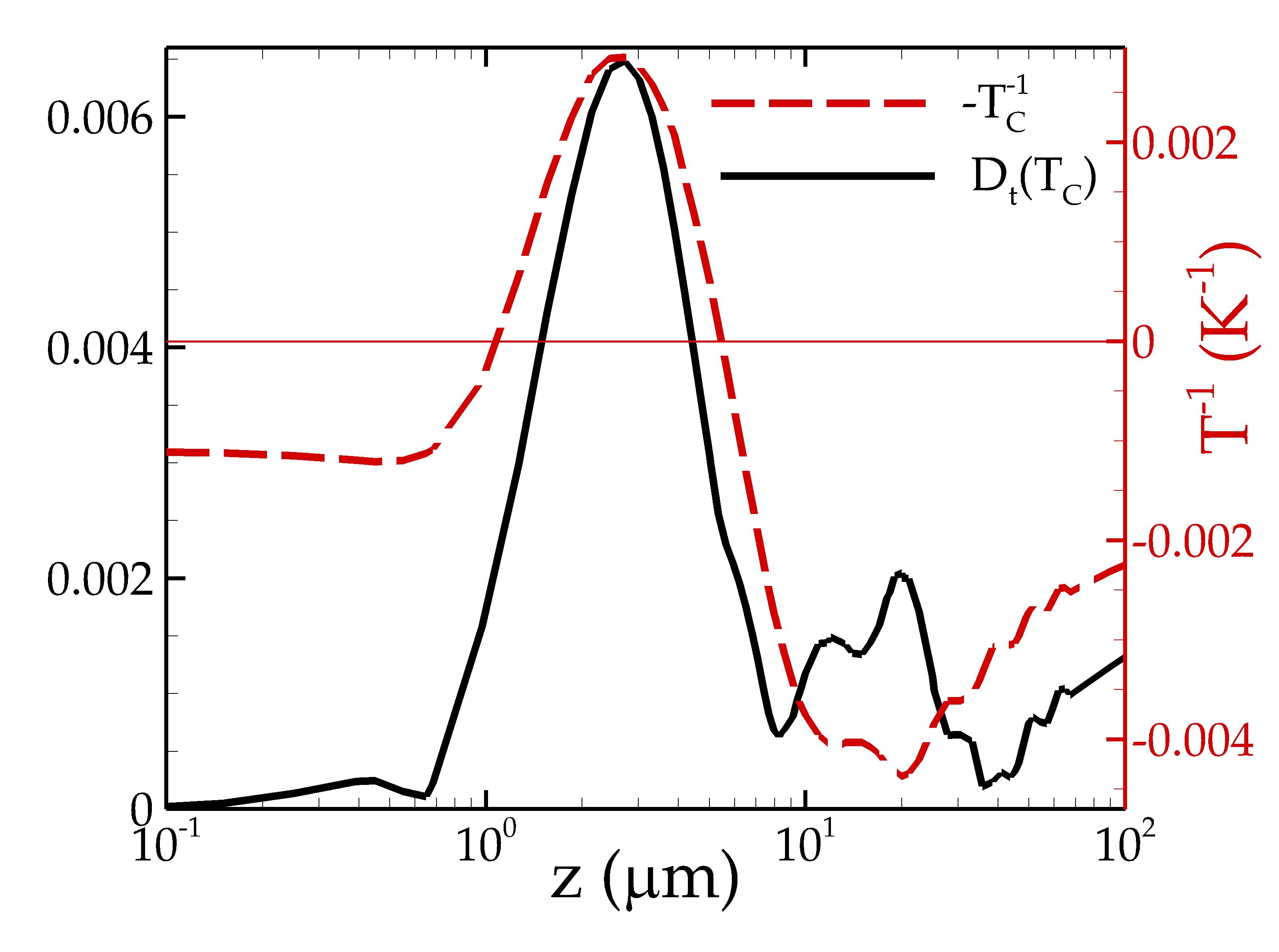}
\end{center}
\caption{Left vertical scale: trace distance $D_\mathrm{t}(T_\mathrm{C})$ (solid black line) between the qubits state $\rho_\mathrm{q}$ and the closest thermal state $\rho_\mathrm{th}(T_\mathrm{C})$ versus $z$. Right vertical scale: inverse of the collective temperature of the qubits state $-1/T_C$ (red long-dashed line). On this plot, we have computed $D_\mathrm{t}(T_\mathrm{C})$, which is the trace distance between $\rho_\mathrm{q}$ and $\rho_\mathrm{th}(T)$ after minimization over $T$ for each value of $z$. The temperature minimizing this trace distance is $T_\mathrm{C}$.}
\label{Figure5}
\end{figure}

Until now, we have analyzed local temperatures of each qubit in the system. However, for several applications a collective many-qubit thermal state could be needed. Strictly speaking, due to the presence of qubit-qubit correlations, the collective qubit state $\rho_\mathrm{q}$ cannot be in the Gibbs form. A legitimate question is thus: how distinguishable is $\rho_\mathrm{q}$ from a collective thermal state $\rho_{\mathrm{th}}(T)$ at temperature $T$ of the qubit system? To answer this, we employ the quantity known as trace distance $D_\mathrm{t}$ \cite{NielsenBook} (Appendix \ref{correlations}4), which tells us how statistically different the outcome of a measurement on $\rho_\mathrm{q}$ is from the one of same measurement on $\rho_{\mathrm{th}}(T)$. The temperature minimizing such a distance is thus what one can define as collective qubit temperature $T_\mathrm{C}$.

Fig.~\ref{Figure5} shows $D_\mathrm{t}(T_\mathrm{C})=D_\mathrm{t}(\rho_\mathrm{q},\rho_{\mathrm{th}}(T_\mathrm{C}))$ and the quantity $-1/T_\mathrm{C}$. Remarkably, the behavior of the temperature of the thermal state is very similar to the one of the population temperature of a single qubit. In particular, also $T_\mathrm{C}$ can go beyond the interval $[T_W,T_S]$ and reach negative values. The trace distance (i.e. the maximal distinguishing probability) has small values, its maximum being of $0.65\%$ reached when $-1/T_\mathrm{C}$ is maximum. Therefore the collective state of the qubits is almost undistinguishable from $\rho_{\mathrm{th}}(T_\mathrm{C})$. This means that M delivers thermodynamic tasks not only on the qubits individually, but also on the collective state of the qubits system as a whole. The tasks performed by the machine on this global state correspond quite strictly to the ones delivered on single qubits.

\subsection{Scaling with temperature}
The functioning of the machine is based on the fact that the system is in an OTE configuration, namely the temperatures of the slab and of the walls are different ($T_S \neq T_W$). Besides, one of the main features of the machine is its aptitude to perform strong thermodynamic tasks on qubits, i.e. to bring their population temperatures outside the range defined by $T_S$ and $T_W$. Note that the atoms-slab distance $z$ and the two external temperatures $T_S$ and $T_W$ are the only parameters on which one can easily exert a detailed control. It is then natural to wonder what happens to the ability of the machine to heat up or cool down the many-qubit system if one changes the values of $T_S$ and $T_W$ rather than $z$.

\begin{figure}[t!]
\begin{center}
\includegraphics[width=245pt]{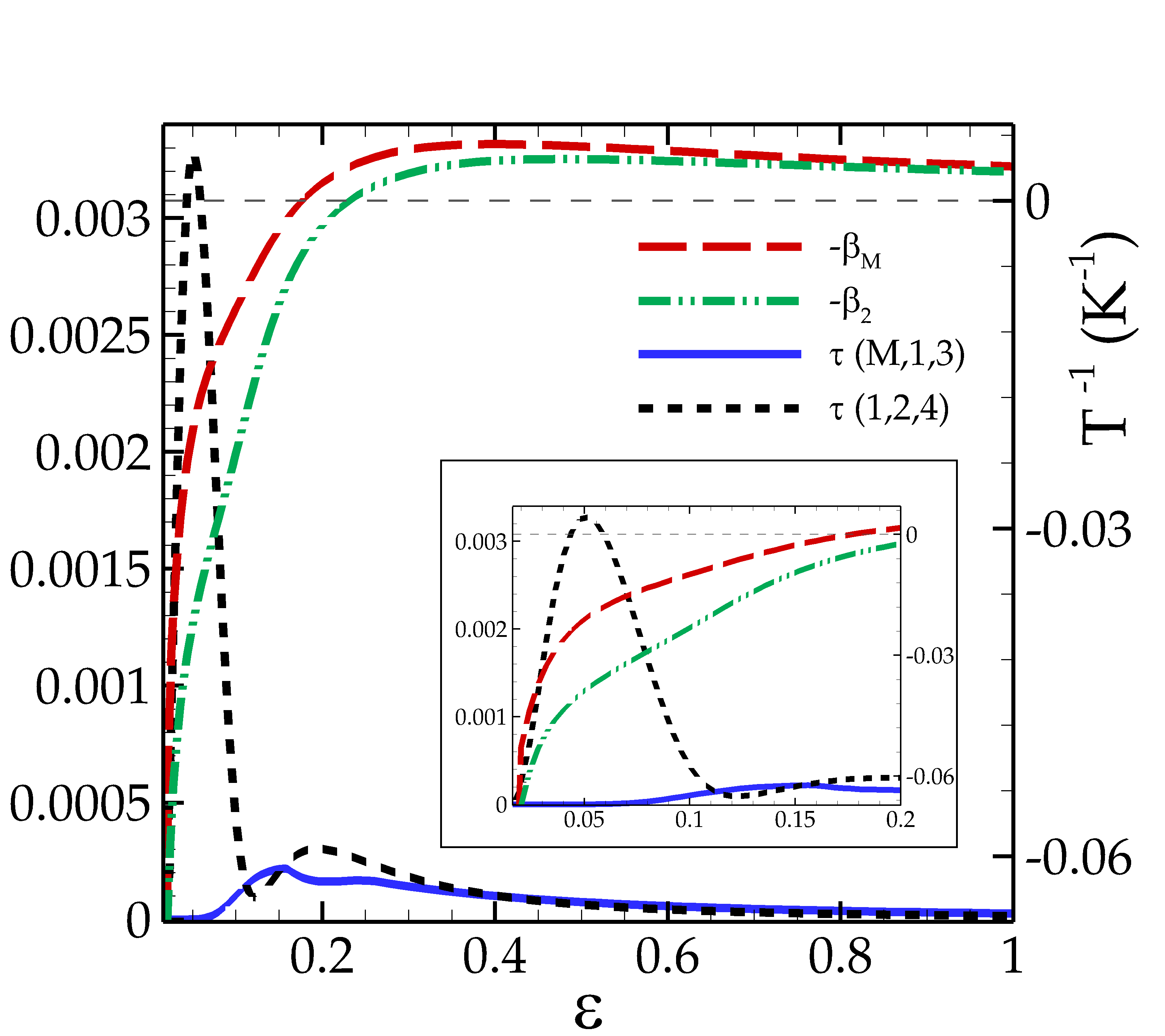}
\end{center}
\caption{Left vertical scale: tripartite correlations for tripartitions $(M,1,3)$ (solid blue line) and $(1,2,4)$ (black dashed line) versus $\varepsilon$. Right vertical scale: inverse of the population temperature of the resonant transition of the machine $-\beta_M$ (red long-dashed line), inverse $-\beta_2$ of the population temperature of qubit 2 (green double-dot-dashed line). The curve $-\beta_2$ is plotted as a representative of the individual population temperatures of the qubits, their behavior being the same. The parameter $\varepsilon$ tunes the externally fixed temperatures as $T_S(\varepsilon)=\varepsilon T_S$ and $T_W(\varepsilon)=\varepsilon T_W$, such that $T_S(1)=900\, $K and $T_W(1)=300\, $K. All the quantities of this plot have been calculated at $z=2.72\, \mu$m. Note that, at $\varepsilon=1$, the configuration is exactly the same as Fig.~\ref{Figure3} at $z=2.72\, \mu$m, at which the machine is heating up the qubits the most effectively (population inversion).}
\label{Figure6}
\end{figure}

To perform this investigation, let us now consider the same configuration of Fig.~\ref{Figure2}, with an atoms-slab distance fixed at $z=2.72\ \mu \mathrm{m}$. This distance corresponds to the one for which the maximum values of $-\beta_{\mathrm{M}}$ and $-\beta_2$ are reached (see Fig.~\ref{Figure3}), i.e. when the action of the machine is strongest. Let us now change the external temperatures through the parameter $\varepsilon \in [0,1]$ as $T_W(\varepsilon)=\varepsilon T_W$ and $T_S(\varepsilon)=\varepsilon T_S$. Fig.~\ref{Figure6} shows the behavior of population temperature of both M and qubit 2 (through $-\beta_{\mathrm{M}}$ and $-\beta_2$) as $\varepsilon$ is tuned.

As shown before, thanks to their highly symmetric configuration, all the qubits tend to distribute the task delivered on them and to equilibrate their population temperatures, such that all their $\theta$ are almost the same. Thus, the behavior of the temperature of qubit 2 we are studying is well representative of the behavior of the rest of the qubits. Fig.~\ref{Figure6} clearly illustrates that, in a large portion of values of $\varepsilon$, (approximately in the range $[0.6,1]$, i.e., $180\,\mathrm{K}\leq T_W\leq300\,\mathrm{K}$ and $540\,\mathrm{K}\leq T_S\leq900\,\mathrm{K}$) the thermodynamics of the qubits-machine system is almost unaffected, highlighting how robust thermal tasks are against a change of $T_W$ and $T_S$. For smaller values of $\varepsilon$, however, $-\beta_{\mathrm{M}}$ and $-\beta_2$ start to decouple. This effect is most clearly seen in the small $\varepsilon$ regime around $\varepsilon\simeq 0.05$ ($T_W\simeq 15\,\mathrm{K}$, $T_S\simeq 45\,\mathrm{K}$), where the difference between $-\beta_{\mathrm{M}}$ and $-\beta_2$ becomes maximal.

This modification of qubits-machine coupling can be interpreted in terms of tripartite correlations in the atomic system. As a matter of fact, as we already pointed out in the previous Section, in order for the two-step thermodynamic task to be effective on all qubits, a balance is needed between correlations in the subsystem $\{\mathrm{M},1,3\}$ and in the subsystems $\{1,2,4\}$ and $\{3,2,4\}$. This is the case for the large $\varepsilon$ interval $[0.6,1]$, where the curves of $\tau(\mathrm{M},1,3)$ and $\tau(1,2,4)$ are almost superimposed. However, for $\varepsilon<0.6$, these two curves are no longer similar and, in particular, for small $\varepsilon$ the qubit-qubit-qubit correlations are much stronger than the M-qubit-qubit ones. This means that one of the two steps of the task cannot be accomplished anymore: the machine is less and less able to affect the qubits state due to the very strong correlations in it. A strong signature of this effect is the fact that the difference $\tau(1,2,4)-\tau(\mathrm{M},1,3)$ is maximal exactly when $-\beta_{\mathrm{M}}$ and $-\beta_2$ are the most different, as shown in the inset of Fig.~\ref{Figure6}. Moreover, in correspondence to this point, no tripartite correlations exist involving the machine ($\tau(\mathrm{M},1,3)=0$), suggesting that qubits and machine are fully decoupled for low-enough temperatures. Notice that the existence of strong entanglement has been predicted in symmetric qubits configurations in exactly this regime of temperatures \cite{Bellomo2015}.
Finally, as thermal equilibrium approaches ($T_W(\varepsilon)=T_S(\varepsilon)$, i.e., when $\varepsilon=0$), all the atomic correlations vanish and all the population temperatures (both of machine and of qubits) collapse on the environmental ones.

\subsection{Scaling with radius}
The previous subsection showed the importance of machine-qubit and qubit-qubit interactions in the delivery of thermodynamic tasks in our system. It is now natural to investigate the dependence of the strength of these interactions between two atoms with respect to the distance separating them. In this subsection, we consider this dependence through the modifications of population temperatures when the radius of the circle along which the qubits are placed changes.

\begin{figure}[t!]
\begin{center}
\includegraphics[width=245pt]{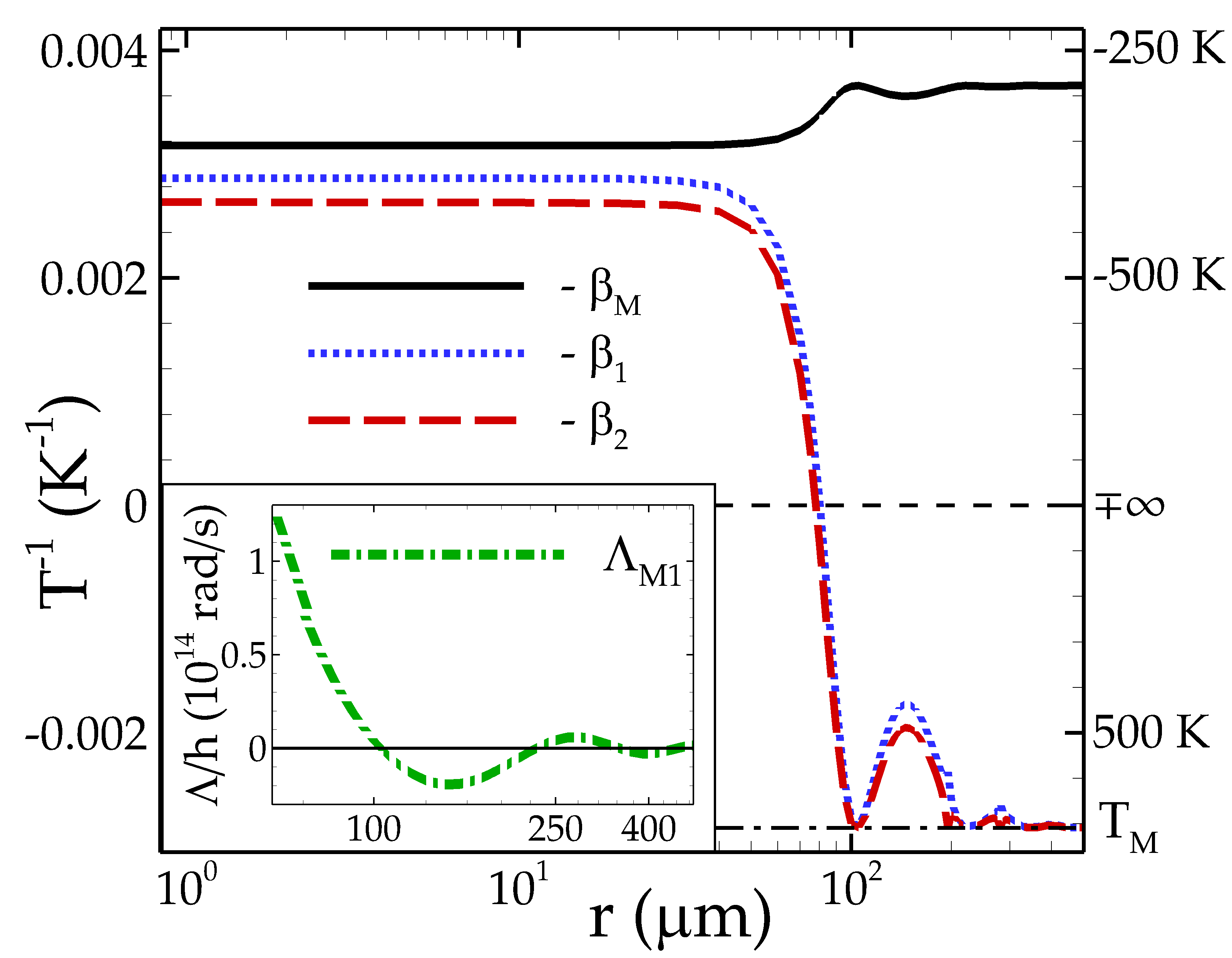}
\end{center}
\caption{Left vertical scale: population temperature of the resonant transition of the machine $-\beta_M$ (solid black line), population temperatures of qubit 1 ($-\beta_1$, blue dotted line) and qubit 2 ($-\beta_2$, red long-dashed line) versus the radius of the circle $r$. Right vertical scale: temperature in correspondence to the left scale. Inset: dipole-dipole interaction amplitude between the machine and qubit 1: $\Lambda_{M1}$ (green dashed-dotted line) versus $r$. All the quantities of this plot have been computed at $z=2.72\, \mu$m, such that at $r=0.833\, \mu$m, the configuration is precisely the same as Fig.~\ref{Figure3} at $z=2.72\, \mu$m.}
\label{Figure7}
\end{figure}

Fig.~\ref{Figure7} reports the changes in population temperatures $-\beta_{\mathrm{M}}$, $-\beta_1$ and $-\beta_2$ for fixed $z=2.72\,\mu$m, when the radius is changed from the value $0.833\,\mu$m to the value $500\,\mu$m. The external temperatures are here again fixed at $T_W=300\,$K and $T_S=900\,$K. The first and most important feature worth stressing here is the fact that, for a remarkably large range of $r$, all the temperatures stay practically constant. Indeed the curves show a plateau up to $r$ as large as $30\,\mu$m. In such a range, all the thermodynamics we have previously described stays unchanged. As such, our previous choice of $r=0.833\,\mu$m is not a limitation, as the same results would have been obtained with any other $r$ in $[0.833, 30]\,\mu$m. Therefore, the functioning of the machine is extremely robust against any uncertainty on the machine-qubits distance.

After such a plateau, a very rapid drop of qubits temperatures is witnessed, together with a slight increase of $-\beta_{\mathrm{M}}$. This marks the transition from the strong to the weak machine-qubit coupling regime. The resonant transition of the machine goes indeed to values very close to the one it would have in the absence of qubits; the same effect can be seen in the qubits temperatures, since they rapidly reach the value of the corresponding environmental temperature, to which they would thermalize in the absence of the machine and which corresponds to the values of the temperatures minimum around $r=10^2\,\mu$m. Finally, some temperature oscillations are seen for larger $r$. These three regimes can be readily explained through the $r$-behavior of the machine-qubit resonant coupling $\Lambda_{\mathrm{M}1}$, shown in the inset of Fig.~\ref{Figure7}. As discussed in the Appendix A (see Eq.~\eqref{lambda0R}), such an interaction has two contributions, one due to the presence of the slab and one induced by the zero-temperature correlations of the field in the absence of matter. This latter is usually dominant, and an analytical expression can be given to it \cite{Bellomo2013b}. This term has two clear limiting behaviors for small and for large atomic separation: when the two atoms are very close (with respect to $c/\omega_\mathrm{q}$), the interaction has a $1/r$ dependence. On the other hand, for $r\gg c/\omega_\mathrm{q}$, this interaction depends on $r$ as a sum of $\sin(r)$ and $\cos(r)$ terms, with a decreasing amplitude.

The plateau of Fig.~\ref{Figure7} is thus a consequence of the rapid growth of $\Lambda_{\mathrm{M}1}$ with decreasing $r$: after a certain threshold, when $\Lambda_{\mathrm{M}1}$ becomes much greater than any other rates involved in the master equation \eqref{METQ}, a saturation effect occurs and all the temperatures become independent of $r$. On the other hand, for $r\sim c/\omega_\mathrm{q}=37\,\mu$m, the transition between these two regimes happens, $\Lambda_{\mathrm{M}1}$ rapidly decreases bringing the temperatures with it and machine and qubits become almost decoupled. Finally, the oscillatory regime of $\Lambda_{\mathrm{M}1}$ produces the residual oscillations of $-\beta_{\mathrm{M}}$, $-\beta_1$ and $-\beta_2$.

This analysis provides also a way to generalize our results to different atomic frequencies: one can be sure that the qubits-machine distance is optimal for thermodynamic tasks as long as it is smaller than the critical value $c/\omega$.

\subsection{Gaussian noise}
Until now we have analyzed the changes induced in the physics of the atoms by parameters on which an external control is easily achievable. A natural problem could however arise if our results were not robust against parameters much harder to control, such as the relative positions of atoms. In preparing realistic systems, indeed, it is not trivial to precisely fix the position of each single constituent. In this subsection, we investigate the robustness of our results against such uncertainty.

To simulate such an uncertainty, we introduce a Gaussian noise on the position of each atom, thus also including the machine. In order to allow for a larger variation of the atomic positions, we use here a larger radius than before, $r=10\,\mu$m, which is however still fully in the plateau zone of Fig.~\ref{Figure7}. The position of each atom is randomly chosen according to a two-dimensional Gaussian distribution, centered on the regular atomic position in Fig.~\ref{Figure2} and with standard deviation on both dimensions fixed at $\sigma=r/10=1\,\mu$m.
Each time the position of a qubit is randomly fixed, its dipole orientation is chosen such that all the dipoles always point toward the machine. The dipole of the resonant transition of the machine, on the other hand, is always kept fixed in the same direction used in the deterministic cases previously studied.

\begin{figure}[h!]
\begin{center}
\includegraphics[width=245pt]{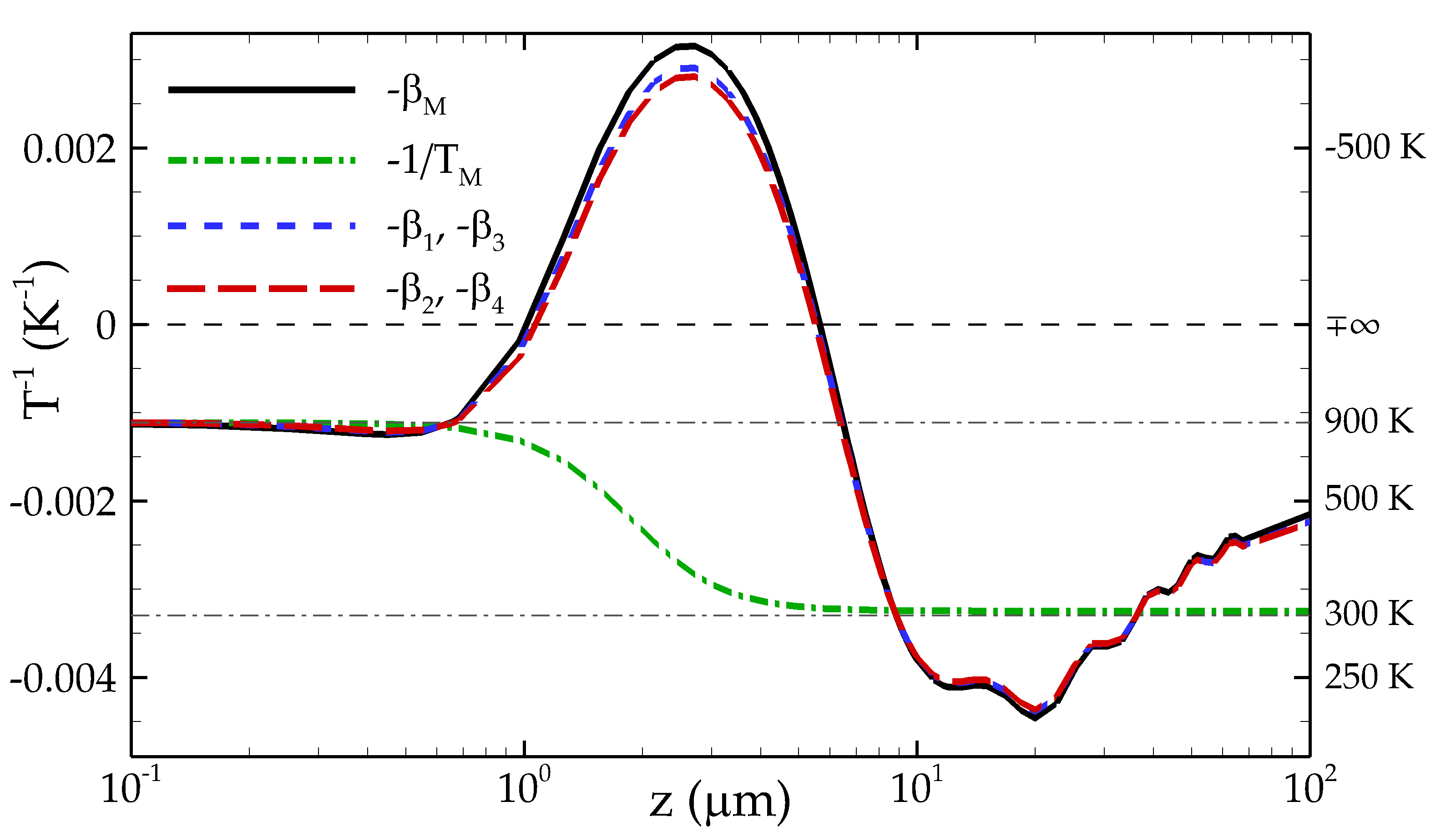}
\end{center}
\caption{
Same quantities and same parameters as Fig.~\ref{Figure3}, except for the radius which has been set here to $r=10\, \mu$m. Note that the difference of radius does not change anything with respect to Fig.~\ref{Figure3} (see Fig.~\ref{Figure7}). These curves have been obtained after averaging over $1000$ realizations. For each realization, the position of each atom has been chosen randomly according to a Gaussian distribution of standard deviation $\sigma=1\, \mu$m, on the two dimensions of the plane containing the atomic system. The dipole of each qubit points toward the machine. The dipole of $M$ points along the direction joining $M$ and the regular position of qubit $1$ (similarly to Fig.~\ref{Figure2}).
}
\label{Figure8}
\end{figure}

For each value of $z$ previously explored we have simulated 1000 random configurations and evaluated, for each of them, all the thermodynamic parameters of interest. We show in Fig.~\ref{Figure8} the averaged temperatures of each qubit and of the resonant machine transition. The $z$-behavior of all the temperatures closely resembles the one shown previously for deterministic positions (see Fig.~\ref{Figure3}), but the maximum of $-\beta$ for each atom is slightly reduced. This is due to the fact that the atomic dipole-dipole coupling is statistically reduced due to the randomness in the relative dipoles orientations stemming from the stochasticity of the atomic positions.

Despite this effect, one sees again that the thermodynamics of the system is very robust also against such a relatively intense random noise: all the qubits still undergo the same thermodynamic tasks as before, in correspondence to the same atoms-slab distances.

\section{Scaling with number of qubits}
\label{scaling}
\begin{figure}[h!]
\begin{center}
\includegraphics[width=245pt]{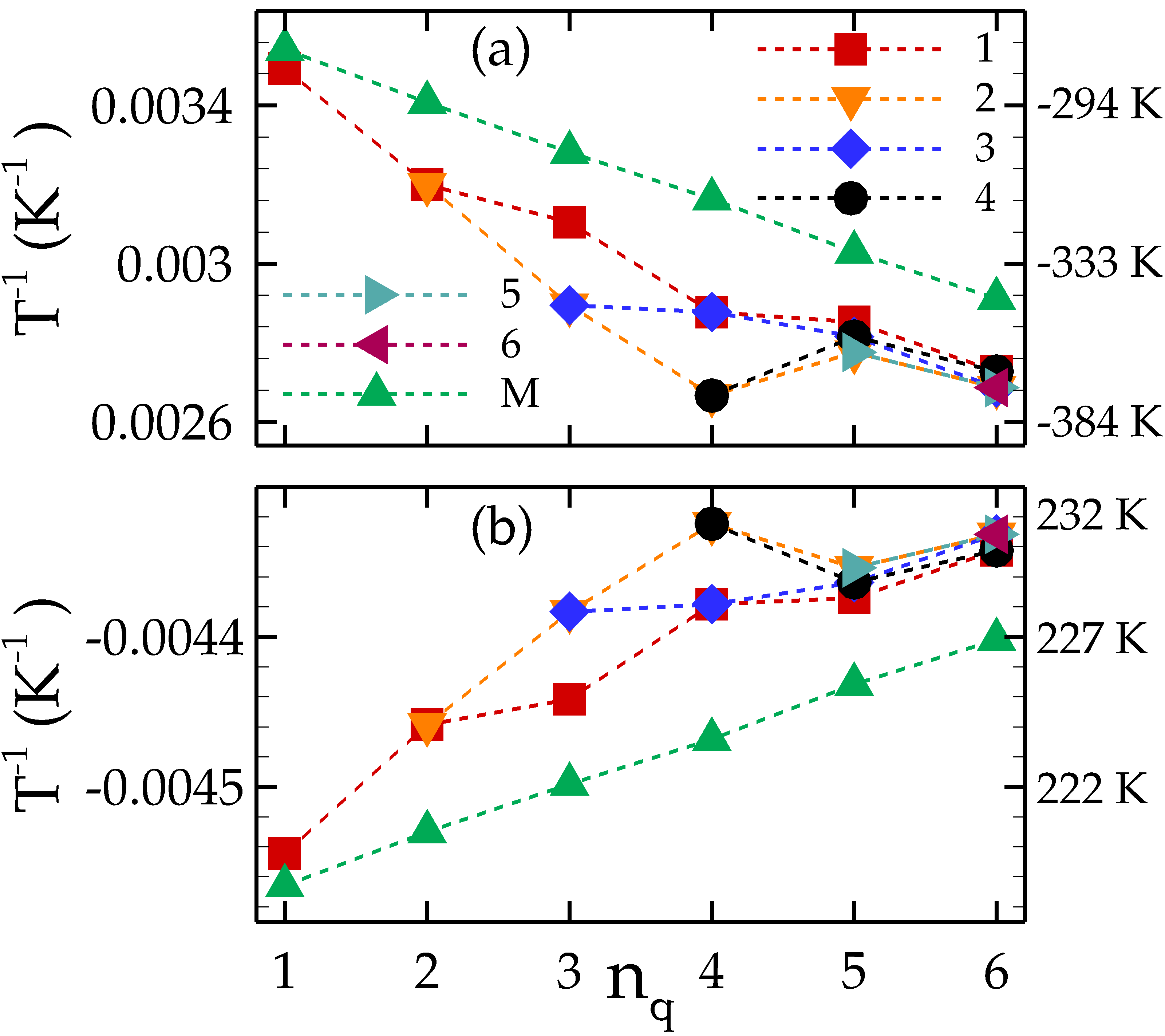}
\end{center}
\caption{On both panels : Left vertical scale: inverse of the population temperatures of each atom versus the number of qubits $n_\mathrm{q}$. Right vertical scale: temperatures in correspondence to the left scale. Panel ($a$) [resp. ($b$)]: maximum (resp. minimum) of the temperatures with respect to the parameter $z \in [0.1,100]\, \mu$m. The other parameters are the same as the ones of Fig.~\ref{Figure3}.
}
\label{Figure9}
\end{figure}

Finally, in this Section we study the scaling of our results with the number of qubits $n_{\mathrm{q}}$, always distributing them regularly along a circle of radius $r=0.833\,\mu$m centered on M. We have fixed $n_{\mathrm{q}}=4$ in all the previous sections as it represents a particularly interesting situation of a two-step task, where the role of correlations is clearer. However, as one sees in Fig.~\ref{Figure9}, similar thermodynamic effects are achieved also with different number of qubits. Fig.~\ref{Figure9}a shows the scaling of the maximum in $z$ of $-\beta$ for all the qubits and for $M$ (i.e., the maximal population inversion induced in each configuration), while Fig.~\ref{Figure9}b shows the minimum of $-\beta$ (maximal refrigeration).

For both of these tasks, two things are worth stressing. First of all, the extremal temperatures of M scale linearly with $n_{\mathrm{q}}$, suggesting that each additional qubit extracts from M or delivers into M (directly or indirectly) the same amount of heat as the qubits already present.

Second, the extremal temperatures of the qubits do not follow such a linear scaling, but rather tend to group together based on the symmetry of the qubits configuration and on the parity of $n_{\mathrm{q}}$: for even qubits numbers, the temperatures collapse to two possible values only, as happens in the case of 4 qubits discussed throughout this paper. On the contrary, for odd qubits numbers, the temperatures tend only partially to group together, there always being an isolated atom at some temperature different from the rest (as easily visible in the case of $n_{\mathrm{q}}=3$ and $n_{\mathrm{q}}=5$). This suggests a collective mechanism of redistribution of the heat exchanged with the machine: pairs of atoms have coupled temperatures if their position along the circle is symmetric with respect to the line joining M and qubit 1. Indeed, the couple $\{\mathrm{M},1\}$ is a privileged one, having always collinear dipoles independently on $n_{\mathrm{q}}$: this also explains why $-\beta_1$ is always the closest one to $-\beta_{\mathrm{M}}$.

Since in this case the radius is constant, adding more and more atoms implies that the qubits are closer and closer to each other, thus increasing their mutual coupling: this has the effect of reducing the difference between their temperatures. Thus, as indeed shown in Fig.~\ref{Figure9}, all the qubits temperatures tend to the same value as the qubits number is increased.

\begin{figure}[h!]
\begin{center}
\includegraphics[width=245pt]{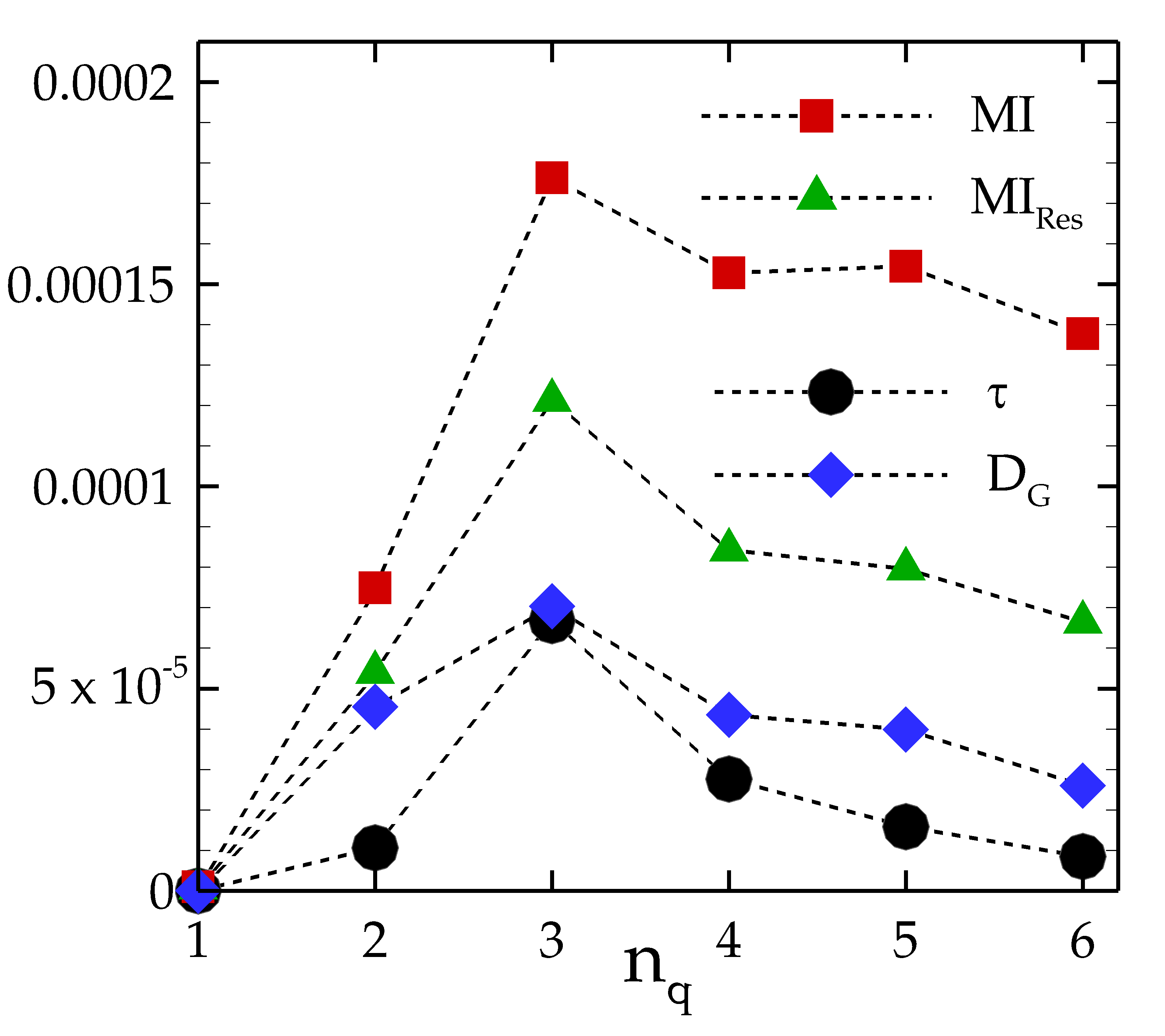}
\end{center}
\caption{Maximum of respectively mutual information MI (red squares), rescaled mutual information $\mathrm{MI}_\mathrm{Res}$ (green triangles), geometric quantum discord (blue diamonds) and tripartite total correlations (black dots) versus the number of qubits $n_\mathrm{q}$. The maximization of each quantifier has been performed for each $n_\mathrm{q}$ over every possible subsystems on which it is defined (e.g. on every tripartitions for tripartite correlations) and with respect to the parameter $z \in [0.1,100]\, \mu$m.
}
\label{Figure10}
\end{figure}

As a further investigation on the $n_{\mathrm{q}}$-dependence shown by physical quantities in our system, we study in Fig.~\ref{Figure10} the scaling with $n_{\mathrm{q}}$ of the three correlation quantifiers employed in the previous analyses: the mutual information MI, the quantum discord $D_\mathrm{G}$ and the total tripartite correlations $\tau$. These three quantities have been maximized, for each $n_{\mathrm{q}}$, over both $z$ and over every possible bi- or tripartition of relevance for the related quantity: every possible bipartition in the atomic system for MI, every possible tripartition for $\tau$ and every possible bipartition of the form $2\times d_B$ (i.e., with one isolated qubit) for the quantum discord, as the analytic formula we employ in its evaluation is valid only under this condition \cite{Spehner2014}.

The scaling behavior of the mutual information rescaled to its theoretical maximum as $\mathrm{MI}_{\mathrm{Res}}=\mathrm{MI}/\max(\mathrm{MI})$ is also shown: as commented in Appendix \ref{correlations}, the theoretical maximal value $\max(\mathrm{MI})$ of MI for a bipartition of dimension $d_A\times d_B$ is $2\ln \big(\min(d_A,d_B)\big)$, since MI quantifies the amount of information stored under the form of correlations between subsystems $A$ and $B$. Since the Hilbert space dimension grows as $2^{n_{\mathrm{q}}}$, more qubits allow more ``memory space" to store information. The scaling of MI provides therefore information on the interplay between the growing Hilbert space dimension and the more and more diluted interactions between subparts. On the other hand, $\mathrm{MI}_{\mathrm{Res}}$ singles out only the $n_{\mathrm{q}}$-scaling of interaction-induced correlations, by providing the relative amount of information with respect to its theoretical maximum. It is interesting to note that the scaling of $\mathrm{MI}_{\mathrm{Res}}$ is very similar to the scaling of the quantum discord, i.e., of the other quantity whose values are normalized in the interval $[0,1]$, independently on the Hilbert space dimension.

Finally, a more technical remark: despite its definition in Eq.~\eqref{discord} of Appendix \ref{correlations}, one should not look here at the difference between MI and discord as measuring some classical correlations: indeed, the geometric measure of discord employed here \cite{Spehner2014} is based on the so-called Bures distance (a legitimate metric in the state space), whereas MI employs an entropic distance as (pseudo)metric. No numerical comparison is therefore possible.

\section{Conclusions}\label{concl}
In this paper we have studied the functioning of a three-level atom as an absorption thermal machine acting on a many-qubit system. This configuration stands in between the two cases studied until now of thermal absorption tasks being delivered, respectively, on macroscopic objects or on single qubits. The extension to the many-qubit case provides here the first step to the application of quantum absorption tasks (i.e., based on quantum features) to realistic scenarios and to systems of applicative interest.
Moreover, it represents a fundamental advancement in the understanding of the role of correlations in the thermodynamics of multipartite quantum systems.

We have indeed demonstrated that thermodynamic tasks can be delivered even if the machine Hilbert space is much smaller than the one of the target body, thanks to the fact that inner correlations in the target body relay and distribute the task on all its parts. This is possible thanks to a realistic and rich out-of-thermal-equilibrium configuration of a single electromagnetic field which acts as a reservoir for the machine functioning. The thermodynamics is therefore based on the interaction with a single, non-equilibrium steady environment.

We have demonstrated the possibility of a detailed control over local temperatures of each qubit in the multipartite system, and over the collective state of all of them. Such a state, despite formally not being under a Gibbs form, is experimentally almost undistinguishable from it as we have shown by the use of the so-called trace distance. Our configuration is therefore able to achieve both a global task on a large quantum system and the same local task on all of its constituents, without the need of either a large thermal machine or of many elementary ones.

We have found a strict correspondence between the change in local and global temperatures induced by the machine and the correlations both inside the target body and between it and the machine itself.

This correspondence, and its consequence of strong and stable thermodynamic tasks, survive over a broad range of temperatures externally fixed to maintain the electromagnetic field in its non-equilibrium steady state.

In addition, we have studied our system under practically realistic conditions, introducing uncertainty on atomic positions and dipole orientation, and tuning the machine-body distance. We have shown that all the tasks delivered by the machine are remarkably robust against these parameters, paving the way for a direct experimental realization of them.

As a suggestion for possible experimental implementations, the role of the qubits could be played by GaAs or InAs quantum dots \cite{Komiyama2000,Kammerer2005,Wasserman2009}, or by the rotational energy levels of water molecules \cite{Dowling1971}. Another possibility is the exploitation of hyperfine structure of atoms, such as oxygen \cite{DeNatale1993}. On the other hand, it has been shown \cite{Gelbwaser2015} that mercure and hydrogen could be good candidates for the realization of a three-level atomic system needed for the machine. Surface array traps have already been used to place atoms above the surface of a material and thus could be exploited to control the positions of the emitters of our system with respect to the sapphire slab \cite{Bermudez2013,Whitlock2009}.

Our results provide for the first time a simple and realistic configuration to have thermodynamic tasks on many-body quantum systems. Remarkably, quantum features in the machine-body interaction and classical correlations inside the body can cooperate in order to achieve absorption thermodynamics with a single elementary quantum machine and a single non-equilibrium environment as its reservoir.

Our findings might be relevant in the recently emerging field of micro- and nanoscopic biosensing, as well as for the local control of many-qubit state during quantum-information tasks \cite{Mitchison2015}.

\section*{ACKNOWLEDGEMENTS}
Authors acknowledge financial support from the Julian Schwinger Foundation.

\begin{appendix}
\section{Rates of the master equation}\label{rates}
Each term contributing to the master equation \eqref{METQ}, whose expressions are given in Eqs.~\eqref{HL}-\eqref{DTQQ}, involves either dipole-dipole interaction strengths $\Lambda_{n\mathrm{M}}$ and $\Lambda_{nm}$, or dissipation rates $\Gamma_n^{\pm}$, $\Gamma_{\mathrm{M}}^{\pm}$, $\Gamma_{n\mathrm{M}}^{\pm}$ and $\Gamma_{nm}^{\pm}$. All of these quantities are obtained within the rotating wave approximation and under the Markovian limit, and are functions of all the system parameters, including properties of the OTE electromagnetic field \cite{Bellomo2013b}. Introducing the vacuum spontaneous emission rate $\gamma^{(i)}_0(\omega)=|\mathbf{d_i}|^2\omega^3/(3\hbar \pi \varepsilon_0 c^3)$ for the transition $i$ (which could be both a qubit or a machine transition) with dipole moment $\mathbf{d}_i$ and frequency $\omega$, the expressions for the single-transition dissipative rates are
\begin{eqnarray}
\frac{\Gamma^+_i(\omega)}{\gamma_0^{(i)}(\omega)}&=&\big[1+n(\omega,T_W)\big]\alpha_W^{(i)}(\omega)\nonumber \\
&+&\big[1+n(\omega,T_S)\big]\alpha_S^{(i)}(\omega),\label{locgp}\\
\frac{\Gamma^-_i(\omega)}{\gamma_0^{(i)}(\omega)}&=&n(\omega,T_W)\alpha_W^{(i)}(\omega)^*+n(\omega,T_S)\alpha_S^{(i)}(\omega)^*,\label{locgm}
\end{eqnarray}
whereas the non-local collective dissipative rates for two transitions $i$ and $j$ of frequency $\omega_\mathrm{q}$ (which can either both be qubit transitions or one of them can be the machine transition resonant with qubits) are given by
\begin{eqnarray}
\frac{\Gamma^+_{ij}(\omega_\mathrm{q})}{\sqrt{\gamma^{(i)}_0(\omega_\mathrm{q})\gamma^{(j)}_0(\omega_\mathrm{q})}}&=&\big[1+n(\omega_\mathrm{q},T_W)\big]\alpha_W^{(ij)}(\omega_\mathrm{q})\nonumber \\
&+&\big[1+n(\omega_\mathrm{q},T_S)\big]\alpha_S^{(ij)}(\omega_\mathrm{q}),\\
\frac{\Gamma^-_{ij}(\omega_\mathrm{q})}{\sqrt{\gamma^{(i)}_0(\omega_\mathrm{q})\gamma^{(j)}_0(\omega_\mathrm{q})}}&=&n(\omega_\mathrm{q},T_W)\alpha_W^{(ij)}(\omega_\mathrm{q})^*\nonumber \\
&+&n(\omega_\mathrm{q},T_S)\alpha_S^{(ij)}(\omega_\mathrm{q})^*.
\end{eqnarray}
The functions $\alpha_W^{(i)}(\omega)$, $\alpha_S^{(i)}(\omega)$, $\alpha_W^{(ij)}(\omega_\mathrm{q})$ and $\alpha_S^{(ij)}(\omega_\mathrm{q})$ depend on the geometrical configuration of the atomic system through their distance $z$ from the slab and each atom-atom distance, and on the geometrical and dielectric properties of the slab. In their explicit expression, not given here for the sake of brevity (the interested reader is referred to Eq.~(33) of \cite{Bellomo2013b} for all the details, where however the factor $\pi$ in the fraction in front of the integrals has to be removed), the dielectric function of the slab material is involved in characterizing the transmission and reflection coefficient of the slab itself. These coefficients come into play when calculating the self-correlation functions of the OTE electromagnetic field, it being given by four contributions: the field coming directly from the walls (under the form of blackbody thermal radiation), the field emitted by the slab, and the two contributions of walls field either reflected by or transmitted through the slab.

Finally, the dipole-dipole interaction strength $\Lambda_{ij}(\omega_\mathrm{q})$, coupling only pairs $(i,j)$ of resonant transitions, has the expression
\begin{equation}\label{lambda0R}
\Lambda_{ij}(\omega_\mathrm{q})=\Lambda_0^{(ij)}(\omega_\mathrm{q})+\sqrt{\gamma^{(i)}_0(\omega_\mathrm{q})\gamma^{(j)}_0(\omega_\mathrm{q})}K_{ij}(\omega_\mathrm{q}),
\end{equation}
where $\Lambda_0^{(ij)}(\omega)$ is the standard free contribution in the absence of matter (slab) stemming from the zero-point correlations of the field, and $K_{ij}(\omega)$ is the reflected contribution which takes again into account the dielectric properties of the slab through its scattering terms. The explicit expressions of $\Lambda_0^{(ij)}$ and $K_{ij}$ can be found respectively in Eqs.~(39) and (C.10) of \cite{Bellomo2013b}, where the vector $\tilde{\mathbf{r}}$ has to be replaced with $\hat{\mathbf{r}}=(\mathbf{R}-\mathbf{R}')/|\mathbf{R}-\mathbf{R}'|$.

\section{Correlations and other measures}\label{correlations}
Here we briefly introduce and discuss the measures of correlations employed in the main text when analyzing the thermodynamics of the atomic steady state. We will give here only few pieces of information, the interested readers being referred to the more specialized literature cited in each subsection.
\subsection{Mutual Information: a measure of total bipartite correlations}
When studying any interaction between two subparts $A$ and $B$ of a multipartite system, the natural question arises about how correlated these two subparts are. This question can be answered by means of the well-known mutual information $\mathrm{MI}(A:B)$ \cite{Cerf1997,Vedral2002}, an entropic measure of shared information. It measures the amount of total bipartite correlations (i.e., quantum \textit{plus} classical bipartite correlations) and it is defined as
\begin{equation}
\mathrm{MI}(A:B)=S(\rho_A)+S(\rho_B)-S(\rho_{AB}),
\end{equation}
where $S(\rho)=-\mathrm{tr}\left(\rho\ln\rho\right)$ is the von Neumann entropy of the quantum state $\rho$. This parameter quantifies the difference between information (as measured by entropy) one has about a composite system $AB$ if only knowledge about the two subparts' state $\rho_A$ and $\rho_B$ is available and the one at disposal by knowing the total state $\rho_{AB}$.

Clearly, if $A$ and $B$ are not correlated, the knowledge of the reduced states equals the knowledge of the composite state and $\mathrm{MI}(A:B)=0$. On the other hand, if $\rho_{AB}$ is a pure state maximally entangled, for which $S(\rho_{AB})=0$, then $A$ and $B$ are maximally correlated. In this case, if $d_A < d_B$, $d_{A(B)}$ being the dimension of the Hilbert space of $A$ ($B$), one has (stemming from the Schmidt decomposition of $\rho_{AB}$) $\mathrm{tr}_{\mathrm{A}}(\rho_\mathrm{AB})=\tfrac{1}{d_A} \mathrm{diag}\big(\mathbb{I}_{d_A},0_{d_B-d_A}\big)$  and $\mathrm{tr}_\mathrm{B}(\rho_\mathrm{AB})=\tfrac{1}{d_A}\mathbb{I}_{d_A}$, where $\mathrm{diag}\big(\mathbb{I}_{d_A},0_{d_B-d_A}\big)$ is the block diagonal matrix composed of the identity matrix of dimension $d_A$ and the null matrix of dimension $d_B-d_A$. Then, for any composite system $AB$, the maximum of mutual information is $\mathrm{MI}(A:B)=2 \ln \big( \min (d_A,d_B) \big)$.

\subsection{Geometric quantum discord}
Another possible question regarding a bipartite system $(A,B)$ is the amount of purely quantum correlations in its state. One can answer this question by studying the quantity called Quantum Discord \cite{Ollivier2001,Modi2012} between $A$ and $B$ (in this order), whose original expression reads
\begin{equation}\label{discord}
D(A\rightarrow B)=\mathrm{MI}(A:B)-\mathcal{C}(A\rightarrow B),
\end{equation}
having defined $\mathcal{C}(A\rightarrow B)$ as the purely classical correlations between $A$ and $B$ (again, the order is here crucial, it being a non-symmetric measure). Clearly, the difficulty of calculating \eqref{discord} stems from the evaluation of $\mathcal{C}$, which involves in general a complicated optimization over the set of POVMs (positive-operator valued measures) on $A$. To overcome such an obstacle, one can introduce a new related measure of discord $D_\mathrm{G}$ as
\begin{equation}\label{geodiscord}
D_\mathrm{G}(A,B)=g(\rho_{AB},\chi_{AB}),
\end{equation}
where $g$ is any valid metric in the state space and $\chi_{AB}$ is the closest \textit{classical state} (i.e., zero-discord state) to $\rho_{AB}$. The quantity in Eq.~\eqref{geodiscord} is known as geometrical quantum discord \cite{Modi2012,Paula2014}.

The advantage of this definition lies in the fact that an analytic formula for $g(\rho_{AB},\chi_{AB})$ is available for some specific cases. In particular, an expression for it is given in \cite{Spehner2014} for bipartitions $2\times d$, where one of the two subsystems is a qubit and the second one can be seen as a $d$-level quantum system. This formula employs the Bures distance $B$ \cite{NielsenBook} as the reference metric.

\subsection{Tripartite correlations}
Another possible piece of information about the distribution of correlations among the different constituents of a multipartite system comes from the study of total tripartite correlations \cite{Giorgi2011,Maziero2012}. Consider thus a tripartite (sub)system $(A,B,C)$: the amount of total (i.e. classical plus quantum) correlations in it is defined as
\begin{equation}\label{tripartite}
\tau(A,B,C)=\mathrm{MI}_3(A:B:C)-\mu(A,B,C),
\end{equation}
where $\mathrm{MI}_3(A:B:C)=S(\rho_A)+S(\rho_B)+S(\rho_C)-S(\rho_{ABC})$ is the total correlation information on the tripartite state $\rho_{ABC}$ and $\mu(A,B,C)=\max\{\mathrm{MI}(A:B), \mathrm{MI}(A:C), \mathrm{MI}(B:C)\}$. In other words, $\tau$ measures the amount of correlations present in $\rho_{ABC}$ which cannot be explained by considering any possible subsystem of $\{A,B,C\}$.

\subsection{Trace distance}
The trace distance $D_\mathrm{t}(\rho,\sigma)$ between two quantum states $\rho$ and $\sigma$ is defined as \cite{NielsenBook}
\begin{equation}
D_\mathrm{t}(\rho,\sigma)=\frac{1}{2}\mathrm{tr}\sqrt{(\rho-\sigma)^2}.
\end{equation}
It is a metric in the state space, with values always in the interval $[0,1]$. In particular $D_\mathrm{t}(\rho,\sigma)=0$ if and only if $\rho=\sigma$ and $D_\mathrm{t}(\rho,\sigma)=1$ if and only if $\rho$ and $\sigma$ have orthogonal supports (i.e., all their eigenvectors with non-zero eigenvalues are orthogonal).

Among its several useful properties, a very operatively clear meaning can be given to its value: the trace distance between $\rho$ and $\sigma$ gives the probability of distinguishing the two states with a single optimal measurement. In slightly more technical words, suppose that one wants to understand, with a single measurement, whether a system is in the state $\rho$ or in the state $\sigma$, these two density matrices having the same \textit{a priori} probability. It can be demonstrated \cite{NielsenBook} that, when employing the optimal measurement to distinguish $\rho$ and $\sigma$, the probability that the measurement outcome allows to understand the state of the system is
\begin{equation}
P=\frac{1}{2}\left(1+D_\mathrm{t}(\rho,\sigma)\right).
\end{equation}
Indeed, if $\rho$ and $\sigma$ have orthogonal supports (thus $D_\mathrm{t}=1$), a definitive answer with $P=1$ can be obtained, provided one measures the system on one of the two supports of $\rho$ or $\sigma$. On the other hand, the more similar the two states are, the less probable is for them to give different measurement outcomes, and the more $D_\mathrm{t} \simeq 0$, leaving one with the only choice to (almost) randomly guess the state of the system with $P\simeq\frac{1}{2}$.

\section{Generalized second law}\label{App2L}
The study of the second law in our configuration is non-trivial due to the non-equilibrium structure of the environment. In particular, applying the Clausius inequality would require the knowledge of the heat fluxes between each component of the system and real thermal reservoirs, which in our case are the slab at $T_S$ and the walls at $T_W$. Nevertheless, it is not possible to determine if a photon emitted by an atom will end up reaching the slab or the walls. Thus the second law in its standard formulation cannot be properly applied here.

This problem has already been encountered in previous works \cite{Correa2014} and tackled in the context of Markovian dynamics by means of a generalization of the second law, reading
\begin{equation}\label{Entro}
\frac{dS_\text{tot}}{dt}=\sum_i\mathrm{tr}\Big[D_i(\rho)\,\log(\rho^\text{ss}_i)\Big]+\frac{dS(\rho)}{dt}\geqslant0.
\end{equation}
where $S_\text{tot}$ is the total entropy (of the open system and its environment), $S$ is the von Neumann entropy of the quantum system, $\rho^\text{ss}_i$ is the kernel of the $i$-th dissipator and $i$ runs over the set of dissipators. The validity of Eq.~\eqref{Entro} is a direct consequence of the Markovian and linear structure of master equation. Note that in the case of real thermal reservoirs, each $\rho^\text{ss}_i$ has a Gibbs form and thus we are left with the standard Clausius inequality. In the case of non-thermal reservoirs such as ours, Eq.~\eqref{Entro} allows to define a parameter playing effectively the role of temperature in entropic fluxes. This can be found by imposing, if possible, a Gibbs structure to $\rho^\text{ss}_i$. In our specific case, this leads to an expression of the effective temperatures equivalent to Eq.~\eqref{Tn}.

\begin{figure}[h!]
\begin{center}
\includegraphics[width=245pt]{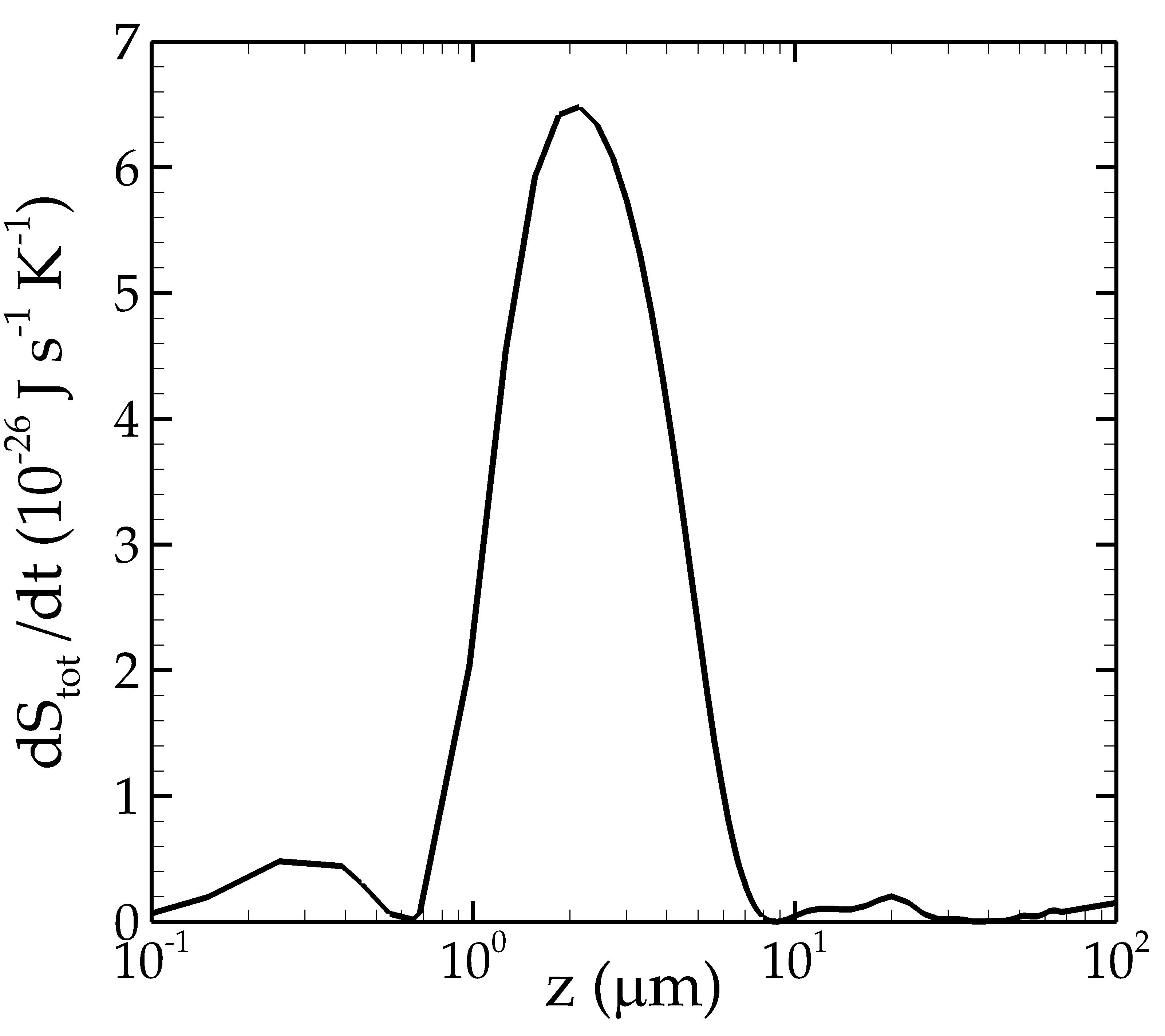}
\end{center}
\caption{Total entropy production in the non-equilibrium steady state of the system as a function of the distance from the slab. The magnitude of the dipoles is $10^{-30}$C.m}
\label{FigureEntropy}
\end{figure}

Equation~\eqref{Entro} allows to follow the total entropy production both during the system dynamics and at stationarity. Figure~\ref{FigureEntropy} reports the value of $dS_\text{tot}/dt$ for the steady state of the system studied in Fig.~\ref{Figure3} as a function of the distance from the slab. As expected, entropy is an increasing function of time for any atoms-slab distance.

One could also wonder whether the heat exchanged by the emitters system with the electromagnetic field has any perceivable effects on the radiative heat transfer between the slab and the walls. Using Eq.~(91) of \cite{Messina2011a}, one can evaluate the heat transfer between slab and wall. For an SiC slab at $T_S=900\,$K in a blackbody radiation at $T_W=300\,$K, such heat transfer turns out to be $\simeq 5.5\times 10^4$\,J\,$\text{s}^{-1}\,\text{m}^{-2}$. This means that, for a slab of $1\,\text{cm}^{2}$, the net heat transfer from the slab to the walls would be of $\simeq 5.5\,$J\,$\text{s}^{-1}$.

Employing now Eq.~\eqref{heatflux} to evaluate all the heat exchanged by the total emitters system and its environment, one finds for $z\mathop{=}2.87\,\mu$m and $r\mathop{=}0.88\,\mu$m a total heat flux of $\mathop{\simeq}-2 \times 10^{-26}$J$\,\text{s}^{-1}$, which is some 26 orders of magnitude smaller than the slab-walls radiative heat transfer. As such, the presence of the atomic system does not affect the energy exchanges between macroscopic objects, and in particular does not reverse the direction of radiative transfer, which still brings heat from the warmer to the colder body, in accordance with the standard macroscopic second law of thermodynamics.

\end{appendix}

\end{document}